\documentclass[aip,jcp,preprint,superscriptaddress]{revtex4-1}

\usepackage{graphicx}

\usepackage{txfonts}
\usepackage{bm}

\usepackage{color}

\begin{document}

\title{Applicability of the Wide-Band Limit in DFT-Based Molecular Transport Calculations}

\author{C. J. O. Verzijl}
\author{J. S. Seldenthuis}
\author{J. M. Thijssen}
\affiliation{Kavli Institute of Nanoscience, Delft University of Technology, 2628 CJ Delft, The Netherlands}

\date{\today}

\begin{abstract}
Transport properties of molecular junctions are notoriously expensive to calculate with \emph{ab initio} methods, 
primarily due to the semi-infinite electrodes. This has led to the introduction of different approximation schemes for the electrodes. For the most popular metals used in experiments, such as gold, the wide-band limit (WBL) is a particularly efficient choice. In this paper we investigate the performance of different WBL schemes relative to more sophisticated approaches including the fully self-consistent non-equilibrium Green's function (NEGF) method. We find reasonably good agreement between all schemes for systems in which the molecule (and not the metal-molecule interface) dominates the transport properties. Moreover, our implementation of the WBL requires negligible computational effort compared to the ground state DFT calculation of a molecular junction. We also present a new approximate but efficient scheme for calculating transport with a finite bias. Provided the voltage drop occurs primarily inside the molecule, this method provides results in reasonable agreement with fully self-consistent calculations.
\end{abstract}

\maketitle

\section{Introduction}

In recent years, approaches to molecular transport based on density-functional theory (DFT) in combination with the non-equilibrium Green's function formalism (NEGF) have received considerable attention in the literature, driven by the rapid progress in experimental work on realizing molecular nano-devices.\cite{Agrait2003,Venkataraman2006,Osorio2007,Quek2007} A number of codes\cite{Taylor2001,Xue2002,Evers2003,Qian2007,Brandbyge2002,Stokbro2003b,Rocha2006} have been developed, including our own implementation within the ADF/BAND quantum chemistry package.\cite{FonsecaGuerra1998,Velde2001,Velde1991,Wiesenekker1991,Verzijl2012}

As the methods have matured, scalability and the computational effort involved in studying progressively larger molecular systems have become important issues. In recent years, the focus of the field has shifted towards systems where the molecule predominantly determines the transport properties, rather than the metal-molecule interface. Examples include systems where quantum interference plays a role.\cite{Ernzerhof2005,Solomon2008,Rincon2009,Markussen2010} This necessitates accurate quantum chemical modeling of the molecule, while the details of the electrodes may be less important to transport. Several approximations exist that trade sophistication for efficiency in the description of the electrodes,\cite{DiVentra2000,DiVentra2001,Ke2004,Rothig2006,Arnold2007} but to the best of our knowledge, there have been no systematic studies of the quality of the results obtained with such approximations.

In this paper we consider methods that treat the molecule at the \emph{ab initio} level of quantum chemical DFT, but vary in the treatment of the electrodes. The most common approach is to use a wide-band limit (WBL) description, which based on the notion that the density of states (DOS) in the electrodes is fairly constant near the Fermi energy (see below). In order to account for the effects of the metal-molecule interface, parts of the electrodes can be included in the DFT calculation, resulting in a so-called ``extended molecule.''\cite{Xue2002,Brandbyge2002,Stokbro2003a}

In the more sophisticated approaches the effect of semi-infinite electrodes is included via self-energies in the self-consistent DFT calculation of the extended molecule. However, this requires the use of a DFT package capable of transport calculations. A common simplifying approach is therefore to perform the DFT calculations on the extended molecule without the self-energies, and only add them afterwards in the calculation of the transport properties. This has the advantage that it can be implemented as a post-processing step which does not require modification of the underlying DFT code (\emph{e.g.}, the ARTAIOS package\cite{ARTAIOS}). Since the open-system character of an extended molecule with self-energies can lead to convergence problems, the closed-system character of this approximation may be advantageous.

In this paper we explore the limits of these approximations, both with and without a bias voltage applied over the molecular junction. We first briefly review the transport formalism which is based on non-equilibrium Green's functions (NEGF). We then introduce the hierarchy of approximations we consider, and compare their computational expense. We evaluate their quality by applying the methods to several junctions containing on phenyl-derivatives, both fully-conjugated and with broken conjugation. Several studies exist in which, for a particular method, the effects of various modeling choices (\emph{e.g.} system size, basis-set, exchange-correlation potential or number of $k$-vectors) on the accuracy have been investigated.\cite{Sen2010,Pontes2011,Ke2004,Ke2005,Ke2007a,Ke2007b,Hoft2007} The present study however, focuses mainly on comparing different methods on the same system.

Although the WBL approximation works well for systems with bulk metal electrodes, it is known to break down for lower-dimensional systems. We will show this explicitly in the case of a monatomic chain, but this can also be an issue in other electrodes such as carbon nanotubes\cite{Guo2006,Feldman2008} (1D) or graphene\cite{Prins2011} (2D). Our results provide an understanding of the applicability and performance of the various approximations for different types of metal-molecule junctions.

We note that the reliability of the transport calculation is limited by the accuracy of the underlying quantum chemistry method. In particular, DFT is well founded for systems in their ground state or, more generally, in thermal equilibrium. In transport, however, we address substantial deviations from the stationary charge distribution, in which case the use of DFT is less well established, in particular because of the shortcomings of the exchange-correlation potential which lacks the derivative discontinuity,\cite{Koentopp2006,Ke2007b} as a result of the neglect of the self-interactions.\cite{Perdew1981,Toher2005} A better alternative is the GW method, which incorporates dynamic response,\cite{Aryasetiawan1998,Thygesen2009} but is prohibitively expensive in terms of computer resources for all but very small transport systems.

\section{Theory}\label{se:Theory}

In transport theory the device is partitioned into electrodes (or leads) and a transport region between them. In the Landauer-B\"uttiker formalism, the current through a junction is given by
\begin{equation}
I=\frac{2e}{h}\int\mathrm{d}\epsilon\left[f_L(\epsilon)-f_R(\epsilon)\right]T(\epsilon),
\end{equation}
where $f_L(\epsilon)$ and $f_R(\epsilon)$ are the Fermi distributions on the left and right electrodes, respectively. The chemical potentials of the leads differ by the bias voltage: $\mu_L-\mu_R=eV_b$. $T(\epsilon)$ is the transmission through the junction, which is given by
\begin{equation}
T(\epsilon)=\mathrm{Tr}\left\{{\bm\Gamma}_L(\epsilon){\bm G}^r(\epsilon){\bm\Gamma}_R(\epsilon){\bm G}^a(\epsilon)\right\},
\end{equation}
where ${\bm \Gamma}_L(\epsilon)$ and ${\bm \Gamma}_R(\epsilon)$ are the couplings to the electrodes and ${\bm G}^r(\epsilon)$ and ${\bm G}^a(\epsilon)$ are the retarded and advanced Green's functions of the transport region, respectively. They are related by ${\bm G}^a(\epsilon)={{\bm G}^r}^\dagger(\epsilon)$. The retarded Green's function is
\begin{equation}\label{eq:Green}
{\bm G}^r(\epsilon)=\left[\epsilon{\bm S}-{\bm H}-{\bm\Sigma}^r_L(\epsilon)-{\bm\Sigma}^r_R(\epsilon)\right]^{-1},
\end{equation}
where $\bm H$ is the Hamiltonian of the transport region, and, in the case of a non-orthogonal basis, $\bm S$ is the overlap matrix of the atomic basis functions. ${\bm\Sigma}^r_L(\epsilon)$ and ${\bm\Sigma}^r_R(\epsilon)$ are the (retarded) self-energies due to the contacts. They can be split into a Hermitian and anti-Hermitian part according to\cite{Jauho1994,Haug1997}
\begin{equation}
{\bm\Sigma}^r_{L,R}(\epsilon)={\bm\Lambda}_{L,R}(\epsilon)-\frac{\mathrm{i}}{2}{\bm\Gamma}_{L,R}(\epsilon).
\end{equation}
For the calculations in this paper, we use a basis in which both ${\bm\Lambda}(\epsilon)$ and ${\bm\Gamma}(\epsilon)$ are real, symmetric matrices. ${\bm\Lambda}(\epsilon)$, induces a shift of the orbital resonances (\emph{i.e.}, the poles of the Green's function), while ${\bm\Gamma}(\epsilon)$ causes a broadening. For certain systems, such as a monatomic chain (see below) in the tight-binding approximation, the self-energies can be calculated analytically.\cite{Newns1969,Anderson1961}

It seems natural to identify the transport region with the molecule, in which case the self-energy is defined on the molecule, specifically near its interface with the metal. However, the self-energy then strongly depends on the contact geometry of the junction and the metal is not allowed to deform its electron density self-consistently near the interface in the presence of the molecule. It is therefore common to include part of the electrodes in the transport region, which then becomes a so-called ``extended molecule,'' and to the self-energy near a metal-metal interface deeper inside the contacts. This makes it possible to use a bulk calculation for the leads to obtain the self-energies, with the added benefit that they only have to be calculated once for a given electrode, irrespective of the molecule.

The imaginary part of the self-energy can be shown to have the following form:\cite{Meir1992,Jauho1994,Haug1997}
\begin{equation}
\Gamma^{L,R}_{nm}(\epsilon)=2\pi \sum_kV^{L,R}_{kn}{V^{L,R}_{km}}^*\delta\left(\epsilon-\epsilon_k\right),
\end{equation}
where $V^{L,R}_{kn}$ couples an electron with momentum state $k$ in the electrode to an atomic orbital $n$ on the molecule. Near the Fermi energy, $V^{L,R}_{kn}$ are generally slowly-varying functions of the momentum $k$.\cite{Meir1992} Moreover, for metals such as gold, the DOS is approximately constant near the Fermi energy\cite{Smith1974,Jepsen1981} (see below). To a first approximation, we can therefore take $\bm\Gamma$ to be independent of $\epsilon$. If we then also neglect the level-shift ${\bm\Lambda}(\epsilon)$, we obtain the wide-band limit (WBL) approximation. This yields a self-energy of the form
\begin{equation}
{\bm\Sigma}^r_{L,R}=-\frac{\mathrm{i}}{2}\Gamma_{L,R}{\bm S},
\end{equation}
where $\bm S$ is again the overlap matrix in the case of a non-orthogonal basis. In the WBL, we have effectively replaced the complexity of the full self-energy by a single parameter $\Gamma_{L,R}$. Although $\Gamma_L\neq\Gamma_R$ in general, in this paper we take them to be the same for simplicity.

Computationally, the main advantage of the WBL is that the eigenspace of the Green's function (Eq.~\ref{eq:Green}) becomes independent of the energy $\epsilon$.\footnote{Technically, this only holds for an orthonormal basis, but it is always possible to transform to such a basis using L{\"o}wdin orthogonalization\cite{Lowdin1950}} We can therefore diagonalize the Green's function before evaluating the transport properties, dramatically reducing the computational cost.

\section{Methods}

\begin{figure}
    \includegraphics[width=6in]{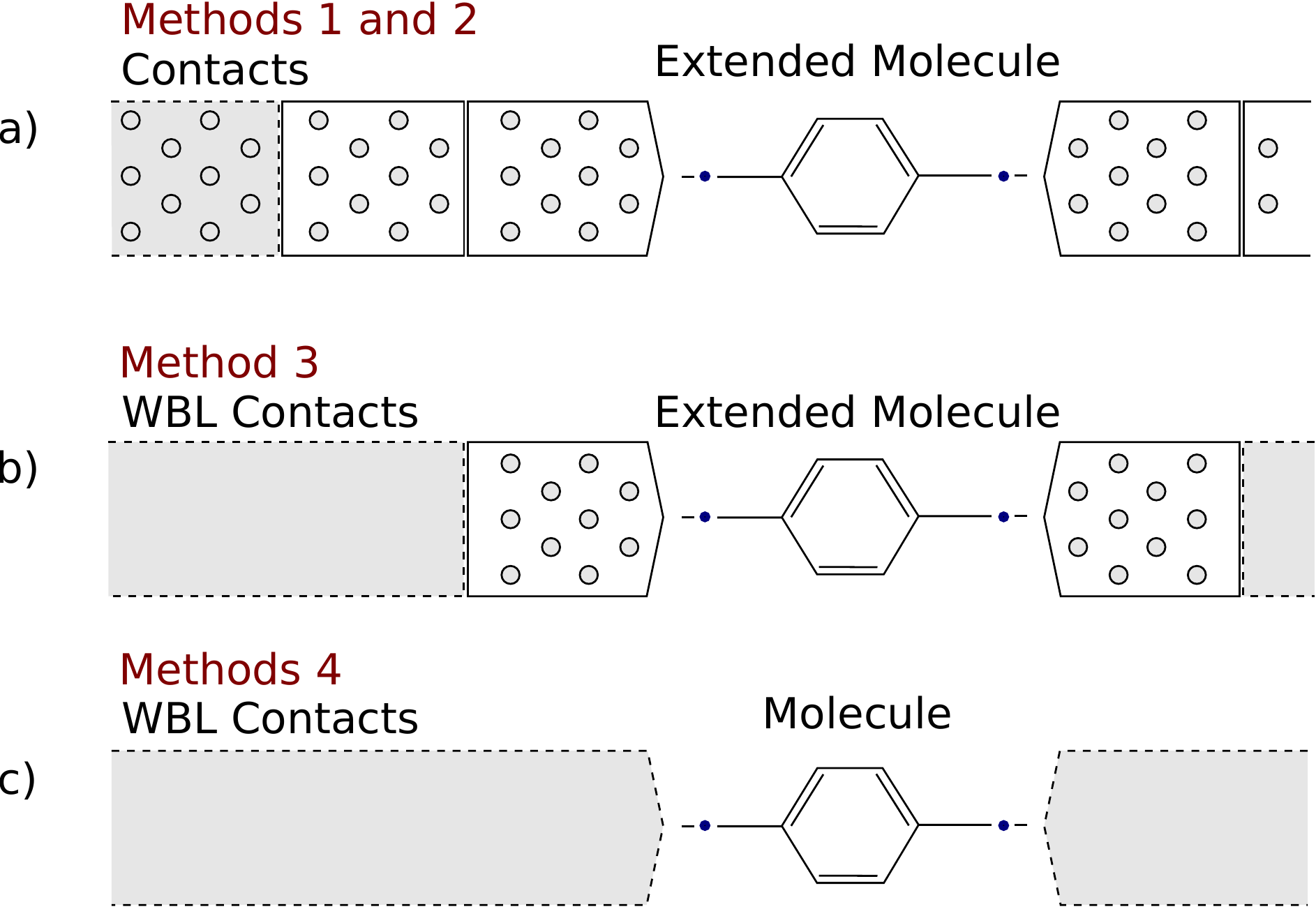}
    \caption{Illustration of geometries corresponding to the four approximations. (a) Full- and Post-SCF methods, with part of the electrodes attached to extended molecule, and full self-energies, corresponding to semi-infinite contacts, behind these. (b) WBL-Metal approach, with WBL self-energies attached to the extended molecule at a metal-metal interface. (c) WBL-Molecule approach with WBL self-energies attached directly to the molecule.}
    \label{fg:Methods}
\end{figure}

\begin{figure}
    \includegraphics{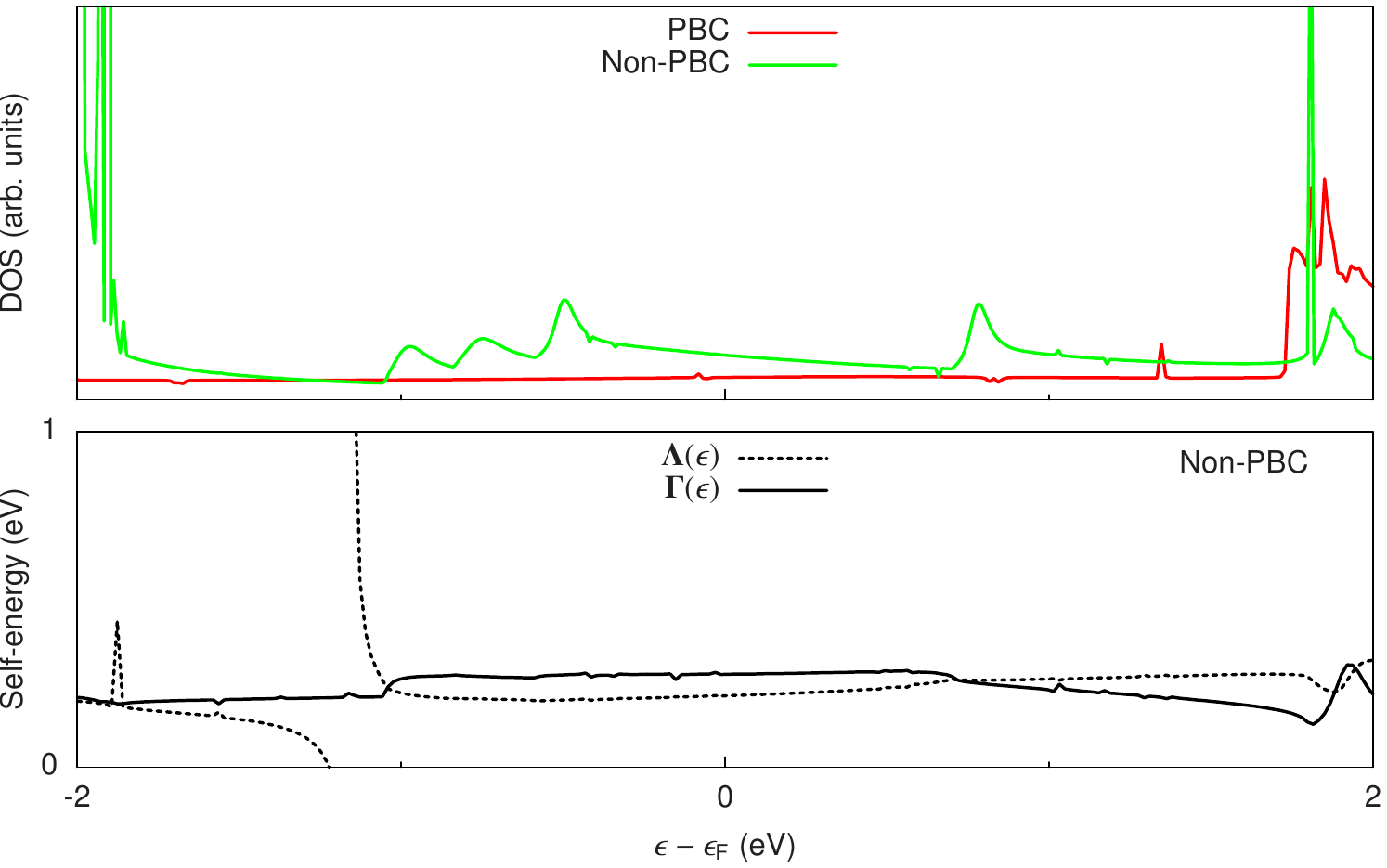}
    \caption{{\bf Top:} DOS of bulk Au calculated with and without periodic boundary conditions. {\bf Bottom:} Magnitude of $\mathrm{Re}\left\{\bm\Sigma\right\}$ and $\mathrm{Im}\left\{\bm\Sigma\right\}$ of bulk Au without PBCs, defined as the average of the diagonal elements of ${\bm\Sigma}(\epsilon)$.}
    \label{fg:Au.dos}
\end{figure}

We evaluate four different approaches for transport calculations (within the NEGF+DFT framework) by comparing their results for a typical metal-molecule-metal junction. In decreasing order of sophistication and computational expense, these are:
\begin{enumerate}
\item\textbf{Full-SCF}: A fully self-consistent transport calculation, where the full self-energies of the metal electrodes are taken into account during the calculation. This corresponds to the steady state of an open system. This approach allows us to include a bias voltage by varying the chemical potential of the electrodes.
\item\textbf{Post-SCF}: A calculation of a closed system, consisting of only the extended molecule, which includes a finite part of the electrodes. The self-energy of the semi-infinite electrodes is then added \emph{after} reaching convergence, in order to enable the calculation of transport properties.
\item\textbf{WBL-Metal}: Similar to Post-SCF, but instead of using the full self-energies in the calculation of the transport properties, the wide-band limit is employed at a metal-metal interface in the electrode.
\item\textbf{WBL-Molecule}: Similar to WBL-Metal, but here the wide-band self-energies are coupled directly to the molecule, and no metal atoms are actually present in the calculation.
\end{enumerate}
The corresponding implications for the geometry in the models are illustrated in Fig.~\ref{fg:Methods}. The Full-SCF method is the reference method to which the other approaches are compared. The difference with Post-SCF gives us an idea of the importance of self-consistency in the calculation. A comparison with WBL-Metal shows the applicability of the wide-band limit for metal electrodes, while a comparison with WBL-Molecule tells us something about the effect of the electrode surface on the molecule.

In the Full-SCF method the bias voltage is applied by varying the chemical potential in the electrodes and simultaneously introducing an electric field inside the junction. Because the calculation is done self-consistently, this yields the correct potential profile in the junction. This cannot be achieved by the other methods, but the effect of the bias voltage can be approximated by introducing an electric field over the molecule. In practice this only works for the isolated molecule, since applying it over an extended molecule leads to convergence difficulties. Finite-bias calculations are therefore only feasible with the Full-SCF and WBL-Molecule methods.

In molecular junctions the magnitude of the field is typically of the order of $1$~V/nm. Since this is much smaller than the internal field of the molecule, the perturbation of the Hamiltonian is effectively linear in the bias voltage. In practice, the Hamiltonian therefore only has to be calculated twice: at zero and at maximum bias. For other voltages the Hamiltonian can then be obtained by linear interpolation. Since the cost of the transport calculation is negligible in the case of WBL-Molecule (see Table~\ref{tab:Expense}), this makes efficient evaluation of the full current-voltage (I-V) characteristics feasible.

Table~\ref{tab:Expense} summarizes the computational expense of the four methods for a typical molecular junction (illustrated in Fig.~\ref{fg:BDT} and discussed in the next section). Timings for the electrode calculations include both DFT and the evaluation of the self-energies. These are not required for the WBL methods, and for the Full-SCF and Post-SCF calculations they only need to be performed once, and so are never a bottleneck. The difference in run time between Full-SCF and Post-SCF is due to the underlying DFT code (see appendix~\ref{se:Details} for details).

There are large differences between the timings of the (extended) molecule calculations. The difference between Full-SCF and Post-SCF is caused by the fact that calculating the electron density of an open system requires integrating the Green's function at every cycle in the DFT calculation.\cite{Verzijl2012} The remainder of the difference is again due to the underlying DFT code. The Post-SCF and WBL-Metal methods use the same calculation of the extended molecule and therefore have the same timings, while the difference with the WBL-Molecule method is due to the absence of the electrodes.

The evaluation of the transport properties is similar for Full-SCF and Post-SCF, resulting in comparable timings. In the WBL methods, on the other hand, the Green's function can be diagonalized independent of energy (see section~\ref{se:Theory}), leading to a speedup of more than two orders of magnitude.

Before applying the methods to a molecular junction, we first consider just gold, the metal generally used for electrodes. Recall that the wide-band limit (section~\ref{se:Theory}) relies on the DOS being nearly independent of energy near the Fermi energy. This is verified in Fig.~\ref{fg:Au.dos}, where we have plotted the DOS and the (average of the trace of the) self-energy of gold as a function of energy. In the case of bulk gold, \emph{i.e.}, with periodic boundary conditions (PBCs), the DOS is indeed constant within $2$~eV of the Fermi energy. For a gold wire (with a cross-section of $3\times 3$ atoms, see appendix~\ref{se:Details}), oscillations reminiscent of Van Hove singularities become visible, due to the essentially one-dimensional nature of the system. However, apart from a single pole at $-1.1$~eV, the self-energy is still approximately constant. This suggests that even for a gold wire, the WBL will be a good approximation, at least at the metal-metal interface.

We will now compare the quality of the different methods by applying them to typical molecular junctions with gold (Au) electrodes, and investigate the effect of the bias voltage on the transmission. For molecular junctions with bulk electrodes both WBL methods give good agreement with the more sophisticated methods. However, in the case of lower-dimensional systems they break down, as we will demonstrate explicitly for a monatomic aluminum (Al) chain.

\begin{table}
    \begin{tabular}{c|ccc}
        Method & \parbox{1.5cm}{Electrodes} & \parbox{1.5cm}{(Extended)\\ Molecule} & \parbox{1.5cm}{Transport\\ Properties} \\[5pt]
        \hline
        Full-SCF  & 36 min & 2.5 hrs & 33 min\\
        Post-SCF  & 23 min & 4 min & 16 min\\
        WBL-Metal & -- & 4 min & 13 s$^*$ \\
        WBL-Molecule & -- & 7 s & 1 s$^*$ \\
    \end{tabular}
    \caption{Comparison of computational expense of the four approaches applied to a typical molecular junction. All calculations have been performed on an 8 core 3 GHz Intel Xeon workstation. $^*$The calculations of the transport properties with the WBL methods have been performed on a single core. }
    \label{tab:Expense}
\end{table}

\subsection{Metal-Molecule-Metal Junctions}

\begin{figure}
    \begin{center}\begin{tabular}{cc}
        \includegraphics[width=3in]{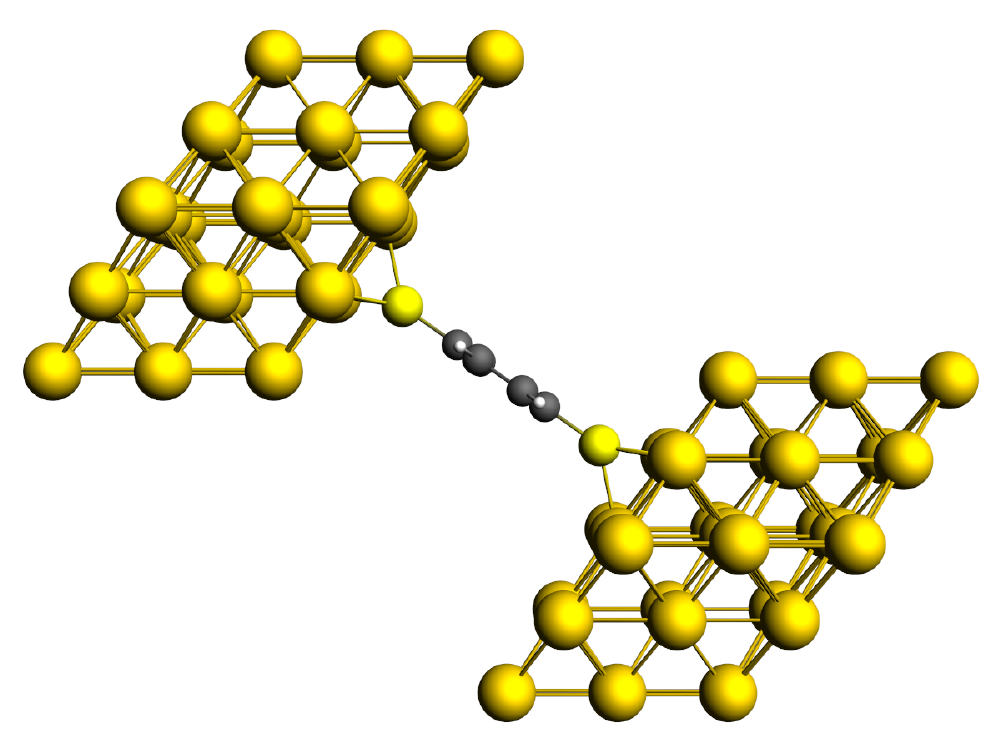} &
        \includegraphics{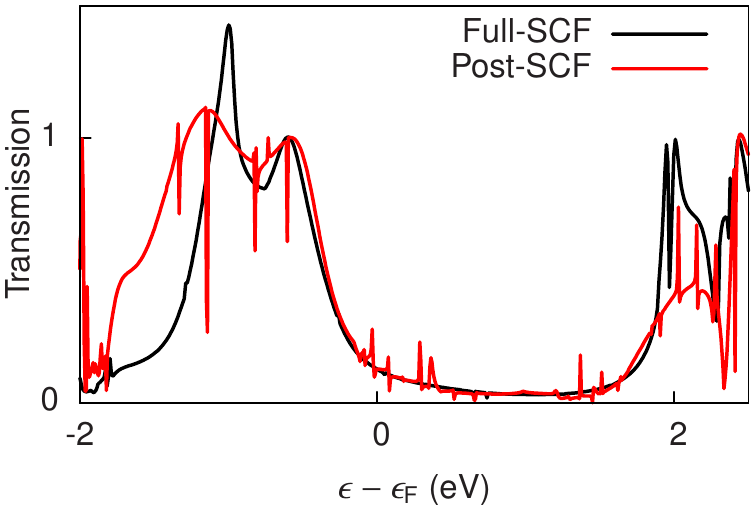} \\
        \includegraphics{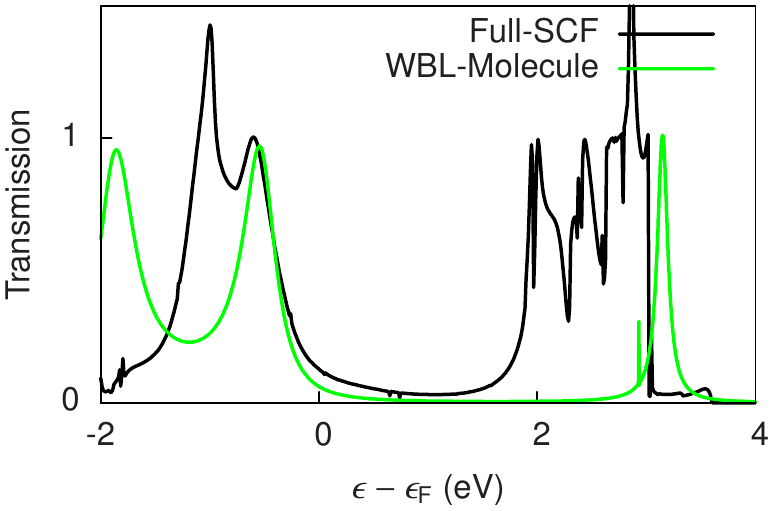} &
        \includegraphics{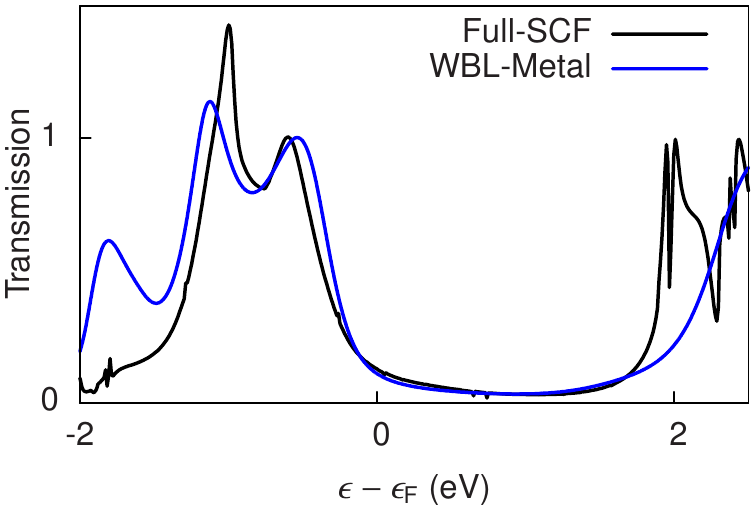}
    \end{tabular}\end{center}
    \caption{Transmission of an Au-BDT-Au junction (top left) for the four different approximations (see text). The Fermi energy of the Full-SCF method is set to $0$~eV. The Post-SCF and WBL-Metal transmissions are shifted by $-0.1$~eV, and the WBL-Molecule transmission by $-0.7$~eV, to align the HOMO peak with that of the Full-SCF calculation. $\Gamma=4$~eV in the WBL-Metal and $\Gamma=0.5$~eV in the WBL-Molecule calculation, respectively. Since the Au electrodes are omitted in WBL-Molecule, the molecule is hydrogen terminated.}
    \label{fg:BDT}
\end{figure}

\begin{figure}
    \begin{center}\begin{tabular}{cc}
        \includegraphics{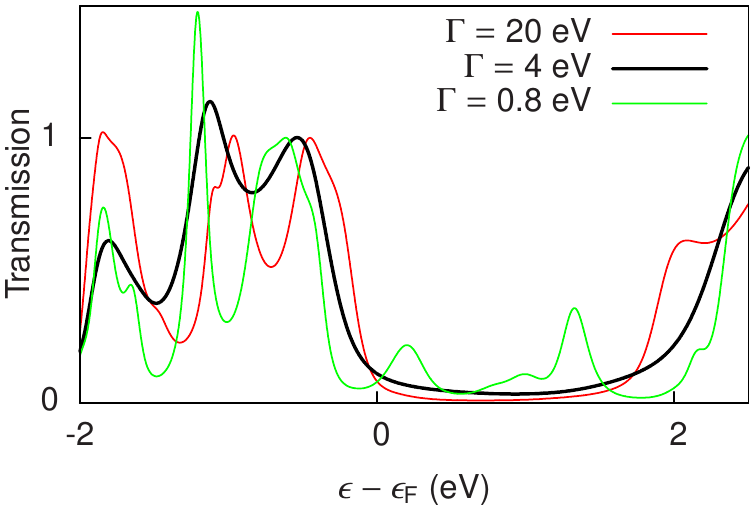} &
        \includegraphics{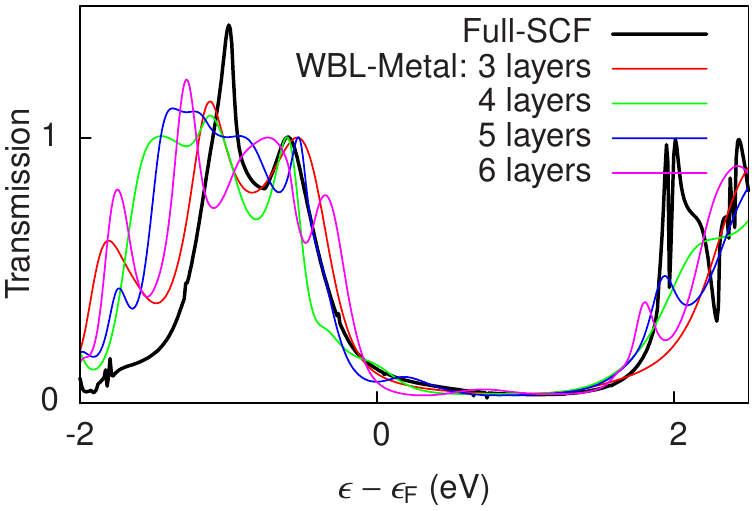} \\
        (a) & (b)
    \end{tabular}\end{center}
    \caption{Transmission of BDT calculated with the WBL-Metal approximation for (a) different values of the coupling strength $\Gamma$ (with 3 layers), and (b) different number of atomic layers of gold in the extended molecule ($\Gamma=4$~eV, \emph{cf.} Fig.~\ref{fg:BDT}).}
    \label{fg:BDT.wbl}
\end{figure}

\begin{figure}
    \begin{center}
        \begin{tabular}{cc}
            \includegraphics[width=3in]{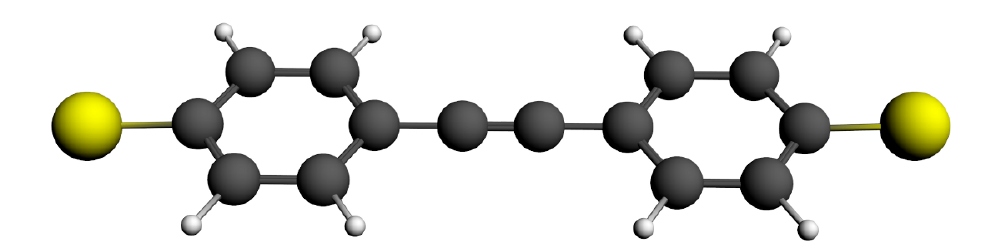} &
            \includegraphics[width=3in]{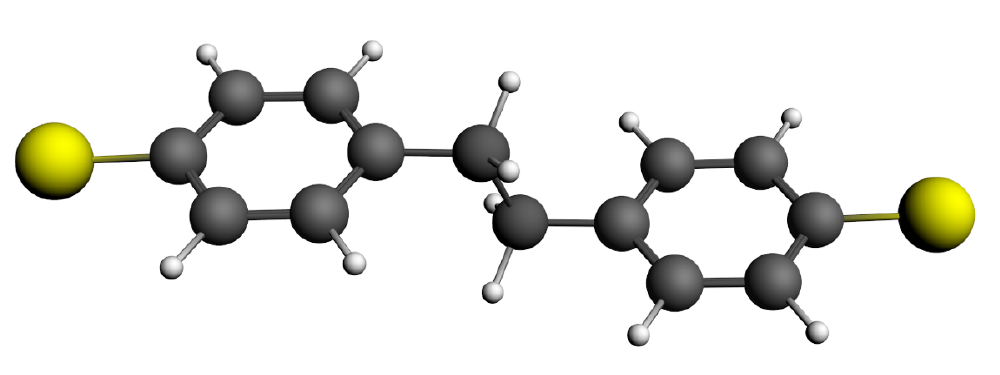}
        \end{tabular}
        \includegraphics{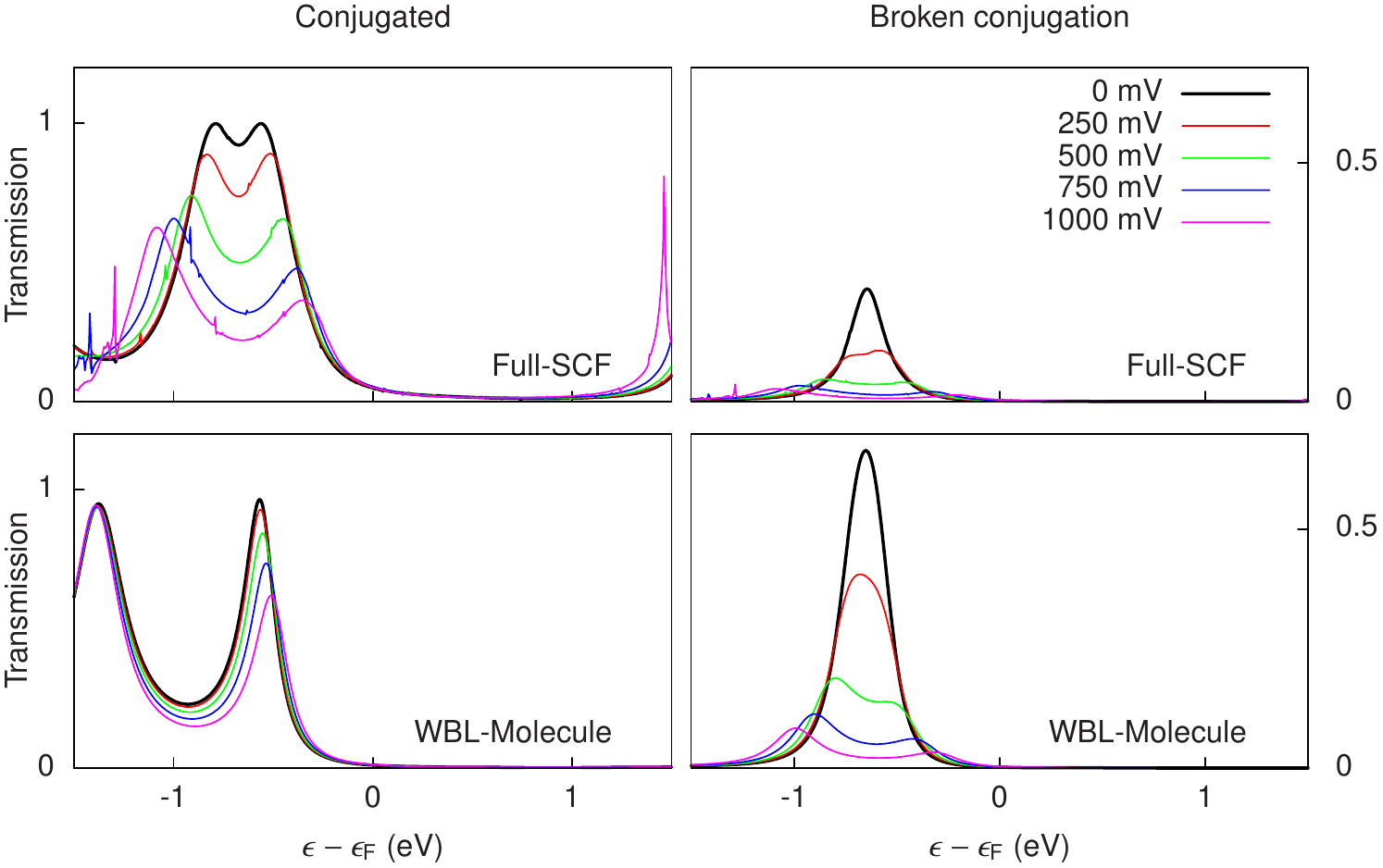}
    \end{center}
    \caption{Transmissions under bias through conjugated and non-conjugated variants of OPE-2 (dithiol). The left panels show transmissions for the conjugated molecule calculated with the Full-SCF (top, with the junction geometry from Fig.~\ref{fg:OPE-2}) and WBL-Molecule (bottom, $\Gamma=0.5$~eV) approaches. The right panels show the same for the broken-conjugation variant. Due to the weak coupling between the left and right halves of the molecule, the HOMO and HOMO-1 appear as a single peak in the transmission at low bias.}
    \label{fg:OPE-2.bias}
\end{figure}

Fig.~\ref{fg:BDT} shows the transmission of a single benzenedithiol (BDT) molecule sandwiched between two Au FCC (111) surfaces. 27 Au atoms are included on either side in the extended molecule (Fig.~\ref{fg:BDT}, see appendix~\ref{se:Details} for details). In order to approximate the geometric configuration of single-molecule break-junction experiments, we do not employ periodic boundary conditions (PBCs), which would mimic a bulk substrate, but rather use needle-like (``non-PBC'') electrodes, or wires\cite{Verzijl2012} (with a $3\times 3$ atom cross-section, see Fig.~\ref{fg:BDT}).

The Full-SCF reference calculation (black lines in the panels of Fig.~\ref{fg:BDT}) shows a broad peak structure slightly below the Fermi energy (set to $0$ eV), corresponding to the HOMO and lower orbitals, and a narrower peak for the LUMO. The HOMO--LUMO gap in the transmission is approximately $2.5$ eV. This transmission is typical for BDT calculations in the literature.\cite{Stokbro2003a,Xue2002,Evers2003,Kondo2006,Verzijl2012}

The red line shows the Post-SCF result, where the HOMO--LUMO gap is the same, but the peaks are slightly broadened with respect to the Full-SCF calculation. Also, the transmission is more ``noisy,'' especially in the HOMO--LUMO gap. The sharp peaks correspond to localized states on the electrodes, caused by the fact that the self-energy is also calculated based on a finite cluster.

The blue line in Fig.~\ref{fg:BDT} shows the transmission calculated with the WBL-Metal approach. The peak structure of the HOMOs and the LUMO corresponds well to the Full-SCF and Post-SCF results, but there is a slight overestimation of the HOMO--LUMO gap.

In the WBL, the magnitude of the imaginary part, which causes a broadening of the peaks, determines the quality of the approximation. This is shown in Fig.~\ref{fg:BDT.wbl}a, where we have plotted the transmission for several values of $\Gamma$. We find that a value of $4$ eV yields the best results. Note that this fit parameter corresponds to the coupling of a metal-metal interface, and is therefore independent of the molecule in the junction.

For $\Gamma=0.4$~eV, extra peaks appear in the HOMO--LUMO gap. As in the Post-SCF case, the peaks correspond to orbitals localized on the electrodes. These orbitals occur in pairs and have a bonding/anti-bonding character, caused by the weak coupling between the electrodes through the molecule. The peaks are unphysical, and disappear when $\Gamma$ becomes larger than the coupling between the electrodes, or when more gold layers are included in the extended molecule.

Adding more gold to the extended molecule will improve the quality of the WBL-Transmission, primarily because it further spatially separates the transport region from the absorbing boundary conditions of the electrodes (due to the imaginary part of the self-energy). The details of the metal-metal interface in the electrodes become progressively less important, as long as the boundary conditions are sufficiently absorbing.\cite{Evers2004,Rothig2006} Fig.~\ref{fg:BDT.wbl}b shows the transmission for 3--6 atomic layers of gold.\footnote{For one or two layers the gold in the extended molecule is not sufficiently bulk-like to attain the correct Fermi energy, and large shifts are necessary to make the transmissions overlap.} Adding gold reduces the width of the cluster of HOMO peaks in the transmission (due to less confinement in the extended molecule), eventually converging to the Full-SCF result. Additionally, a ``shoulder'' becomes visible at $2$~eV, which reduces the apparent HOMO--LUMO gap and leads to a better correspondence with the Full-SCF transmission.

Finally, the bottom left panel of Fig.~\ref{fg:BDT} (green line) shows the WBL-Molecule result. Although the transmission reproduces the double-peak structure of the HOMOs and the sharper peak of the LUMO, the gaps between the orbitals are overestimated by roughly a factor of 2. This is a result of the fact that the calculation is performed on a molecule in gas-phase. This misses hybridization of the molecular orbitals with the electrodes, which reduces confinement of the electrons and is found to significantly reduce the gap in the other approaches. Note also that by omitting the electrodes in WBL-Molecule calculations, there is no clear definition of the Fermi energy; hence the transmission has been shifted to align the HOMO-peak with that of the Full-SCF calculation.

For small molecules, the transmission is dominated by the metal-molecule interface, and WBL-Molecule is a poor approximation. For larger molecules, especially ones where the conductance is small (for example, due to broken conjugation), the transmission is instead dominated by properties of the molecule, and the use of WBL-Molecule becomes appropriate. This is especially clear in the case of finite-bias calculations.

Fig.~\ref{fg:OPE-2.bias} shows the transmission of two oligophenylene-ethynylene (OPE-2) derivatives as a function of bias voltage, calculated with the Full-SCF and WBL-Molecule methods (see Fig.~\ref{fg:OPE-2} for the zero-bias transmissions calculated with the other methods). On the left, the transmission of the fully conjugated OPE-2 molecule is shown. In the Full-SCF calculation two HOMO peaks are visible near the Fermi energy. Upon application of a bias voltage, the peaks move apart and their magnitude decreases. This effect is not reproduced in the WBL-Molecule calculation, where only the HOMO peak shifts and drops in magnitude. As with BDT, the WBL-Molecule approximation overestimates the splitting between the two peaks.

When we break the conjugation (right-hand side of Fig.~\ref{fg:OPE-2.bias}), only one peak is visible close to the Fermi energy at zero bias, which splits into two peaks for higher bias voltages. Moreover, even at zero bias, the transmission is significantly less than one. This effect is reproduced in the WBL-Molecule calculation, although the height of the peak is overestimated.

The difference between the two cases can be understood by considering how the bias voltage is distributed over the junction. For a conjugated molecule, where the $\pi$-electron cloud is easily deformed, resulting in a high polarizability, the voltage drop occurs at the metal-molecule interface. Since this interface is not taken into account in the WBL-Molecule calculation, its results do not agree with the Full-SCF approach. When we break the conjugation, we introduce a barrier, and the voltage drop will occur primarily inside the molecule.\cite{Xue2003} Since the molecule is accurately modeled in the WBL-Molecule approach, the Full-SCF result is qualitatively reproduced. The same effect occurs in the case where amine linkers replace the thiols (see appendix~\ref{se:OPE-2}).

Analysis of the levels involved in transport reveals that the {HOMO and HOMO-1} are composed of bonding and anti-bonding combinations of orbitals located on one of the phenyl rings. In the case of the conjugated OPE-2, the coupling is strong ($388$ meV for WBL-Molecule), and the peaks are split at zero bias. In the case of broken conjugation, the coupling is much weaker ($70$ meV), resulting in the appearance of a single peak at low bias. Additionally the low coupling causes a reduction in the magnitude of the transmission.

\subsection{Monatomic Chains}

\begin{figure}
    \includegraphics{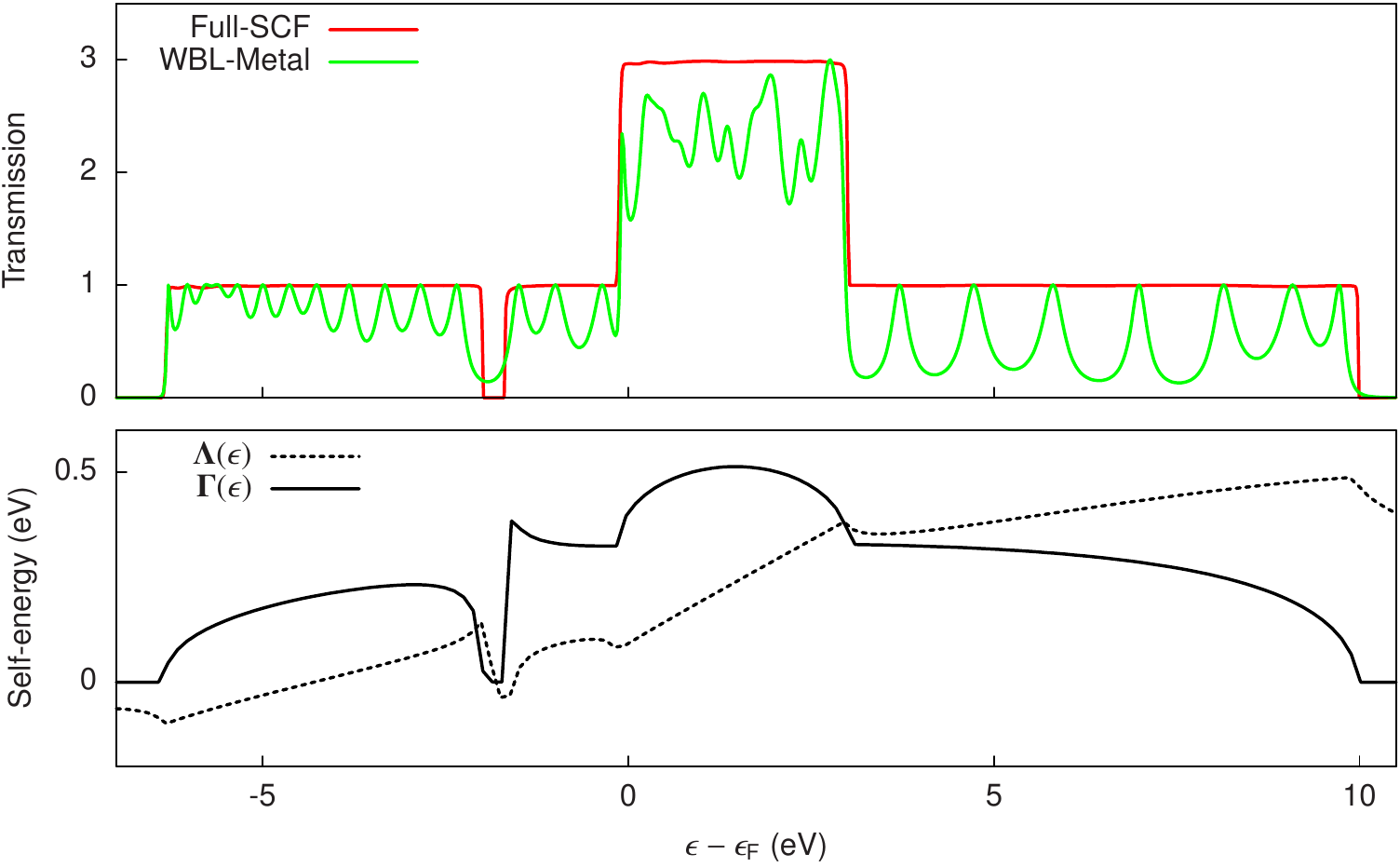}
    \caption{(a) Transmission of a monatomic Al chain, calculated with both the Full-SCF and WBL-Metal ( $\Gamma=0.5$~eV) approaches. (b) Real and imaginary parts of the trace of the self-energy.}
    \label{fg:Al.chain}
\end{figure}

In the case of bulk (3D) contacts the WBL-Metal approximation shows good quantitative agreement with the Full-SCF approach. However, this approximation breaks down for lower-dimensional electrodes. In a typical one-dimensional chain, the DOS is only approximately constant in a narrow range around the Fermi energy, but quickly deviates away from it, culminating in Van Hove singularities near the band edge.\cite{Ashcroft1976} For a monatomic chain in the tight-binding approximation, this can be calculated analytically.\cite{Newns1969,Anderson1961}

Fig.~\ref{fg:Al.chain}a shows the transmission of a monatomic aluminum (Al) chain, calculated with the Full-SCF (red line) and WBL-Metal (green line) approaches. The Full-SCF transmission shows two plateaus, one at a transmission of $1$, with a small dip at $-2$~eV, and one at a transmission of $3$. Although the WBL-Metal approach reproduces the maxima of the transmission and the dip at $-2$~eV, it does not reproduce the plateaus, but it oscillates instead.

This behavior can be understood by considering the self-energy shown in Fig.~\ref{fg:Al.chain}b. For each plateau, $\Lambda(\epsilon)$ has a linear, and $\Gamma(\epsilon)$ a semi-elliptical energy-dependence. Their combined effect is to merge the peaks into a plateau. Although $\Gamma(\epsilon)$ is approximately constant at its maximum, the fact that it drops to zero at the band edge of the chain is responsible for the step-function behavior of the transmission at the band edge. Neither of these effects is captured by the wide-band limit approach, and so the transmission remains a series of broadened peaks.

\section{Conclusions}

In conclusion, we have shown that the wide-band limit approximation gives reasonable to very good results for transport calculations on molecular junctions. In our implementation, the evaluation of the transport properties adds only a few seconds (on a modern workstation) to a standard ground-state DFT calculation of the junction. The cheapest approach consists of adding WBL electrodes to an isolated molecule (WBL-Molecule), the DFT calculation of which takes only a few seconds. Including part of the electrodes in the extended molecule (WBL-Metal) increases the computational cost of the DFT calculation to a few minutes. When the self-energies of the electrodes are included in the transport calculation in a non self-consistent way (Post-SCF), the latter becomes the bottleneck. Due to the open nature of the system, including the self-energies in the DFT calculation as well (Full-SCF) can increase the computation time to several hours.

The Full-SCF approach is the reference method. It has the benefit that it takes the metal-molecule interface into account in a self-consistent way, and allows for the application of a finite bias voltage. The Post-SCF approach quantitatively reproduces the zero-bias transmission, but does not allow for finite bias-voltage calculations. The WBL-Metal approach also gives rather good results for a molecule coupled to three-dimensional electrodes, but it breaks down for lower-dimensional electrodes. Finally, the WBL-Molecule approach qualitatively reproduces the main features of the transmission. However, the estimated energy-gaps can be off by a significant amount due to the omission of interface effects such as chemisorption. In cases where the transmission is dominated by the properties of the molecule, it can reproduce the bias-voltage dependence. Exploiting the linear dependence of the molecular Hamiltonian on the bias voltage, this can be implemented very efficiently, allowing for fast evaluation of the full current-voltage characteristics.

In practice we recommend starting with the WBL-Molecule approach as it is computationally very cheap on any modern workstation. Additionally, lack of hybridization of the orbitals with the electrodes greatly simplifies the analysis of the features in the transmission.\cite{Solomon2008} This is especially important for molecules which are expected to exhibit interference effects.

For quantitative predictions, the WBL-Metal, Post-SCF and Full-SCF approaches give similar results at zero bias. If a dedicated transport code is available, we recommend the Full-SCF approach (which also accounts for the application of a bias voltage); if not, the other approaches can be implemented as post-processing steps after a conventional DFT calculation. However, with WBL-Metal care must be taken when modeling junctions using 1D or 2D electrodes (such as carbon nanotubes\cite{Guo2006,Feldman2008} or graphene\cite{Prins2011}), and more generally for electrodes with a more complicated electronic structure near the Fermi energy.

\begin{acknowledgments}
This research was carried out with financial support from the Dutch Foundation for Fundamental Research on Matter (FOM), and the EU FP7 program under the ``ELFOS'' grant agreement. The authors would also like to thank Gemma Solomon and Kyungwha Park for fruitful discussions about the manuscript.
\end{acknowledgments}

\clearpage
\appendix

\section{Computational Details}\label{se:Details}

All DFT calculations have been performed with the Amsterdam Density Functional (ADF) quantum-chemistry package,\cite{FonsecaGuerra1998,Velde2001,Velde1991,Wiesenekker1991,Verzijl2012} using the LDA exchange-correlation potential.\footnote{Using the PBE GGA potential shifts the transmission slightly, but otherwise leaves the spectrum unchanged.} A single-$\zeta$ (SZ) basis-set was used for the electrodes, and a triple-$\zeta$ polarized (TZP) basis-set was used for the molecule, although we find similar results with a DZP basis.

For the electrodes, we use non-periodic contacts with a FCC (111) surface consisting of $3\times3$ Au atoms (see Fig.~\ref{fg:BDT}). Unless otherwise specified, three atomic layers of gold are included on either side in the extended molecule. The thiolated molecules are situated perpendicular to this surface above a hollow site with a Au--S distance of $2.40$~\AA\ (see Fig.~\ref{fg:BDT}). In the case of the diamines, the molecules are bonded to an adatom on the surface with a Au--N distance of $2.28$~\AA\ (see Fig.~\ref{fg:OPE-2A}), similar to that used by Ning \emph{et al.}\cite{Ning2007} and Quek \emph{et al.}\cite{Quek2007}.

We have implemented the Full-SCF method in the periodic band-structure code BAND, which is a part of the ADF package. At the time of writing, BAND is not as optimized as ADF, leading to the differences in timing in Table~\ref{tab:Expense}. The Post-SCF and WBL-Metal methods have both been implemented in the GREEN module in ADF, while the WBL-Molecule method is a simple Python program.

In the Full- and Post-SCF methods, the self-energies are obtained from a calculation of bulk electrodes. These are modeled as a stack of principal layers, each of which consists of three $3\times3$ atomic layers. In the Post-SCF and WBL-Metal calculations, the extended molecule includes a principal layer on either side of the molecule, for a total of 54 Au atoms. For technical reasons, a Full-SCF calculation contains 90 Au atoms.\cite{Verzijl2012} Finally, in the WBL-Molecule approach, the WBL self-energies are coupled directly to the $p_z$ orbital on the thiols or amines, which our studies indicate to be the dominant charge-injection pathway into the molecular system.

\section{OPE-2}\label{se:OPE-2}

\begin{figure}
    \begin{center}\begin{tabular}{cc}
        \includegraphics[width=3in]{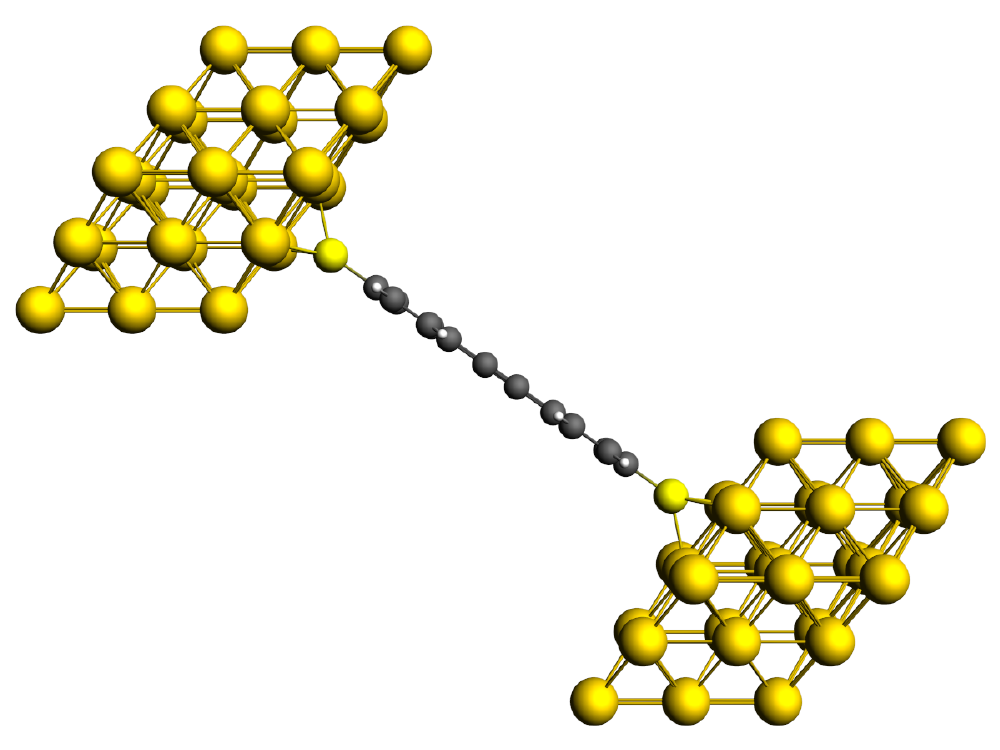} &
        \includegraphics{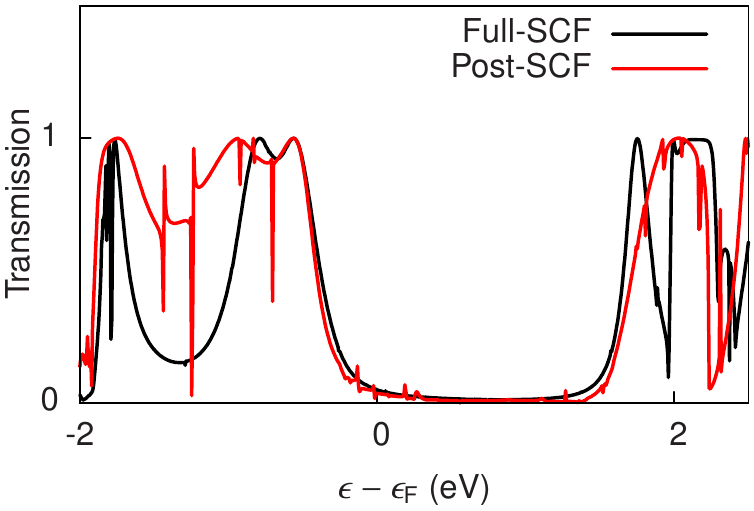} \\
        \includegraphics{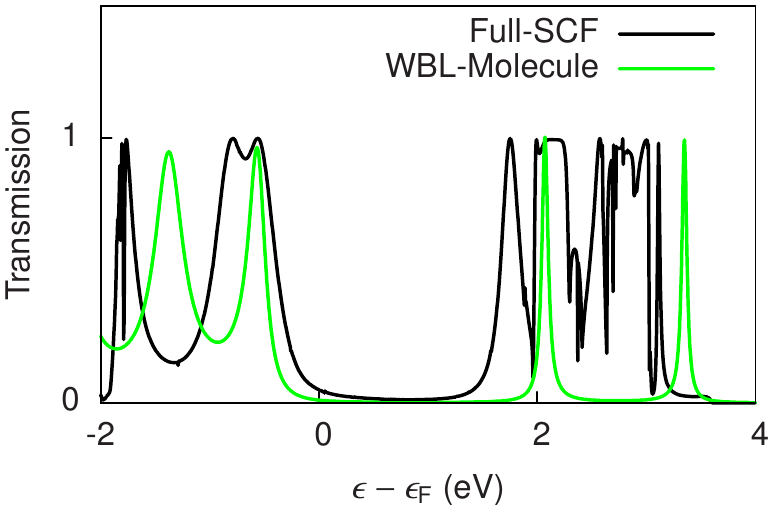} &
        \includegraphics{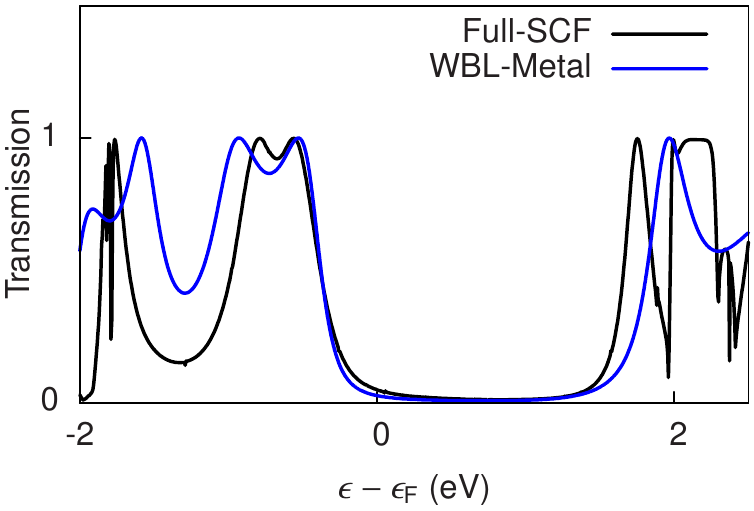}
    \end{tabular}\end{center}
    \caption{Transmission of an OPE-2 (dithiol) junction calculated with all four methods. The Post-SCF and WBL-Metal transmissions are shifted by $-0.2$~eV, and the WBL-Molecule transmission by $-0.9$~eV. The other parameters are the same as in Fig.~\ref{fg:BDT}.}
    \label{fg:OPE-2}
\end{figure}

\begin{figure}
    \begin{center}\begin{tabular}{cc}
        \includegraphics[width=3in]{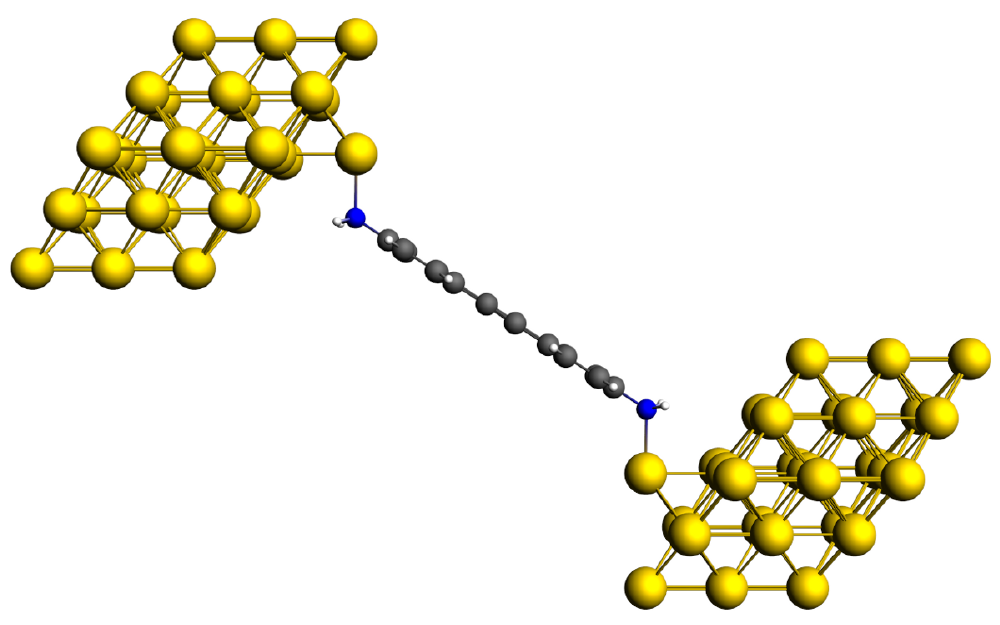} &
        \includegraphics{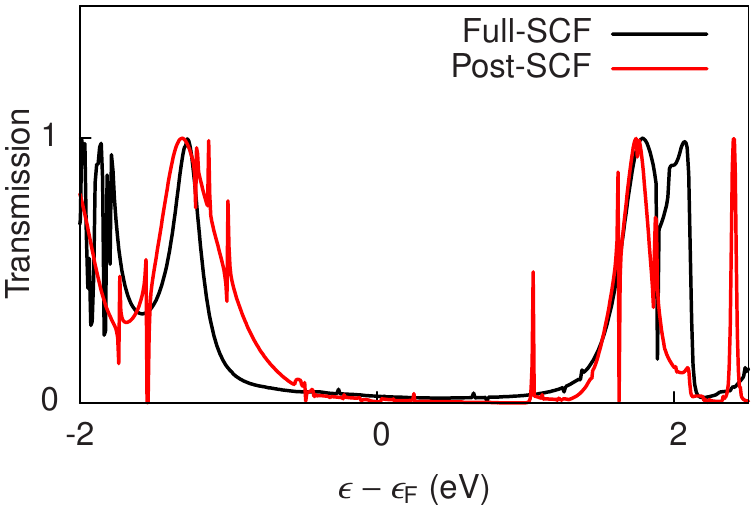} \\
        \includegraphics{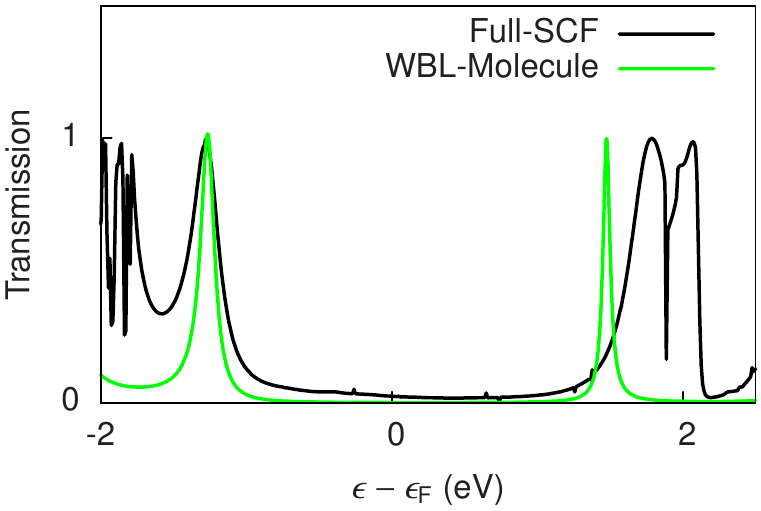} &
        \includegraphics{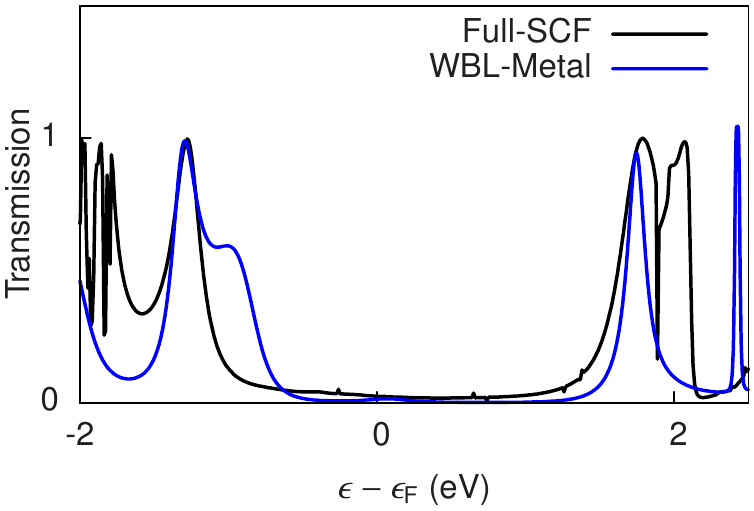}
    \end{tabular}\end{center}
    \caption{Transmission of an OPE-2 (diamine) junction calculated with all four methods. The Post-SCF and WBL-Metal transmissions are shifted by $-0.5$~eV, and the WBL-Molecule transmission by $-2.1$~eV. The other parameters are the same as in Fig.~\ref{fg:BDT}.}
    \label{fg:OPE-2A}
\end{figure}

\begin{figure}
    \begin{center}
        \begin{tabular}{cc}
            \includegraphics[width=3in]{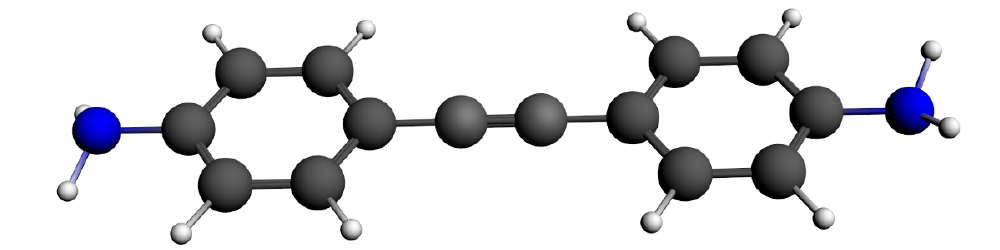} &
            \includegraphics[width=3in]{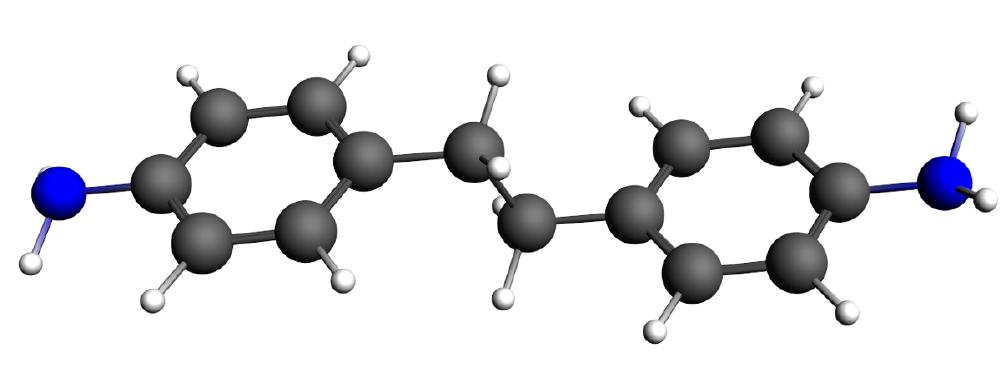}
        \end{tabular}
        \includegraphics{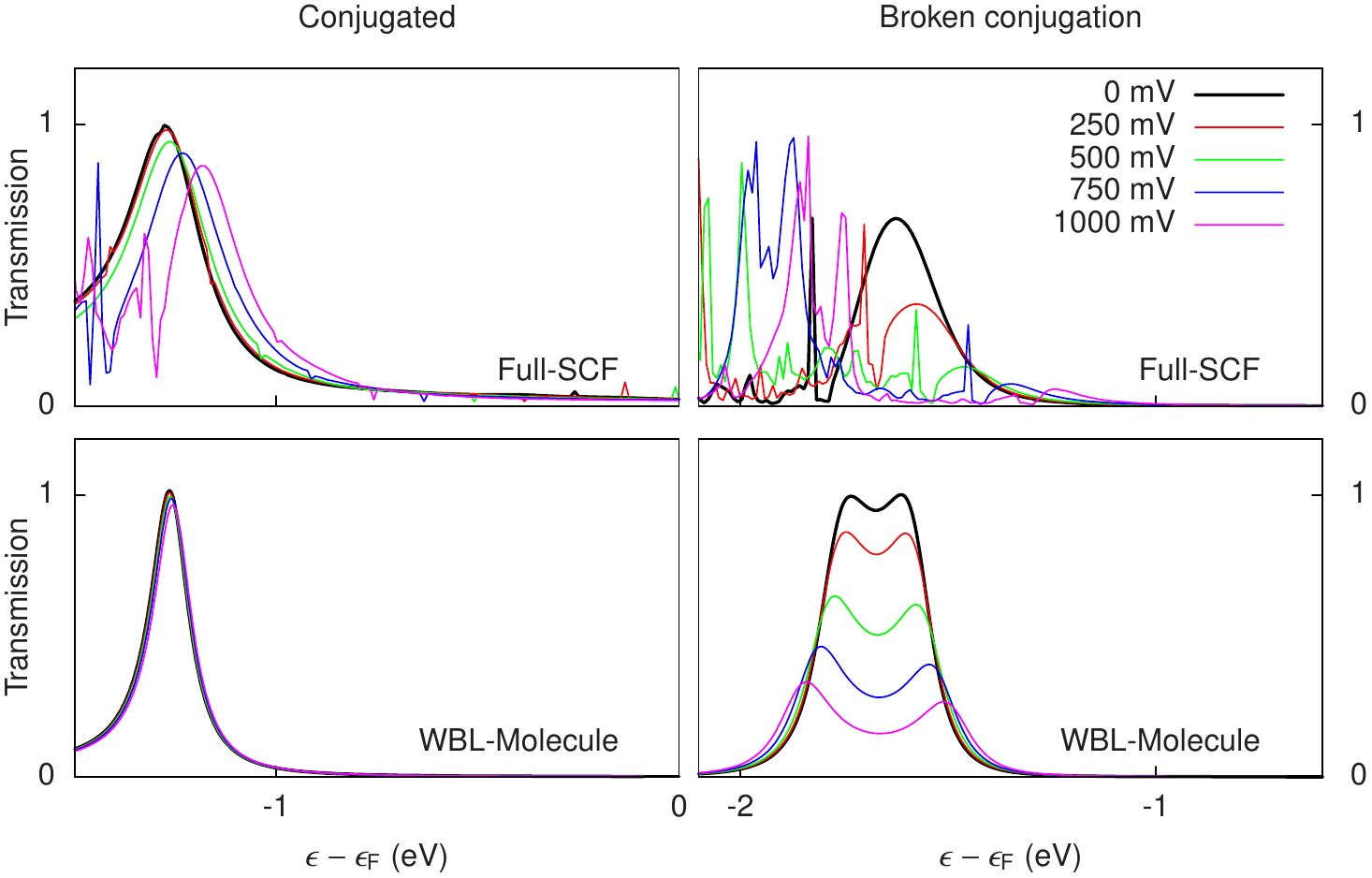}
    \end{center}
    \caption{Transmissions under bias through conjugated and non-conjugated variants of OPE-2 (diamine). As in Fig.~\ref{fg:OPE-2.bias}, the left panels show transmissions for the conjugated molecule calculated with the Full-SCF (top) and WBL-Molecule (bottom, $\Gamma=0.5$~eV) approaches, while the right panels show the same for the broken-conjugation variant. The Full-SCF calculation uses the junction geometry from Fig.~\ref{fg:OPE-2A}.}
    \label{fg:OPE-2A.bias}
\end{figure}

In the main text, we have presented results for molecular junctions with thiol-gold couplings. To illustrate that our results are not exclusively applicable to this particular case, we discuss the OPE molecule with amine-linkers and 
compare the results with those obtained for the thiol-terminated molecule. We show results for both cases, using all 
methods.

Fig.~\ref{fg:OPE-2} shows the zero-bias transmission of OPE-2 (with thiol linkers), calculated with the four different methods. As in the case of BDT (Fig.~\ref{fg:BDT}), the Post-SCF and WBL-Metal transmissions are in good agreement with the Full-SCF result. The splitting between the HOMO peaks is again slightly overestimated. For WBL-Molecule, the HOMO--LUMO gap is closer to the Full-SCF gap than in the case of BDT. This is explained by the fact that the difference in confinement between the isolated and extended molecule is smaller for OPE-2 than for BDT.

Exchanging thiol linkers for amines (Fig.~\ref{fg:OPE-2A}) leads to a larger HOMO--LUMO gap and a single sharp peak in the transmission for the HOMO. This is explained by the weaker coupling of the amine, which binds to a single gold adatom.\cite{Ning2007} Interestingly, both the Post-SCF and WBL-Metal, but not the WBL-Molecule, transmissions show a ``shoulder'' to the right of the HOMO peak. This peak corresponds to a bonding/anti-bonding pair of orbitals on the gold, although the coupling through the molecule is stronger than in the case of Fig.~\ref{fg:BDT.wbl}a. It does, however, decrease for larger values of $\Gamma$.

For the conjugated version of OPE-2 (amine), the bias-dependence of both the Full-SCF and WBL-Molecule transmission (Fig.~\ref{fg:OPE-2A.bias}) is similar to that of OPE-2 (thiol). For the broken-conjugation, the WBL-Molecule transmission follows the same trend as in Fig.~\ref{fg:OPE-2.bias}, although the peak is split even at zero-bias. The rapid oscillations in the Full-SCF transmission below the HOMO peak make it difficult to determine whether
such splitting is present. The HOMO peak itself does follow the same trend as for WBL-Molecule (and Fig.~\ref{fg:OPE-2.bias}); the peak decreases in magnitude and shifts up in energy for higher bias.

%

\clearpage
\begin{thebibliography}{57}%
\makeatletter
\providecommand \@ifxundefined [1]{%
 \@ifx{#1\undefined}
}%
\providecommand \@ifnum [1]{%
 \ifnum #1\expandafter \@firstoftwo
 \else \expandafter \@secondoftwo
 \fi
}%
\providecommand \@ifx [1]{%
 \ifx #1\expandafter \@firstoftwo
 \else \expandafter \@secondoftwo
 \fi
}%
\providecommand \natexlab [1]{#1}%
\providecommand \enquote  [1]{``#1''}%
\providecommand \bibnamefont  [1]{#1}%
\providecommand \bibfnamefont [1]{#1}%
\providecommand \citenamefont [1]{#1}%
\providecommand \href@noop [0]{\@secondoftwo}%
\providecommand \href [0]{\begingroup \@sanitize@url \@href}%
\providecommand \@href[1]{\@@startlink{#1}\@@href}%
\providecommand \@@href[1]{\endgroup#1\@@endlink}%
\providecommand \@sanitize@url [0]{\catcode `\\12\catcode `\$12\catcode
  `\&12\catcode `\#12\catcode `\^12\catcode `\_12\catcode `\%12\relax}%
\providecommand \@@startlink[1]{}%
\providecommand \@@endlink[0]{}%
\providecommand \url  [0]{\begingroup\@sanitize@url \@url }%
\providecommand \@url [1]{\endgroup\@href {#1}{\urlprefix }}%
\providecommand \urlprefix  [0]{URL }%
\providecommand \Eprint [0]{\href }%
\providecommand \doibase [0]{http://dx.doi.org/}%
\providecommand \selectlanguage [0]{\@gobble}%
\providecommand \bibinfo  [0]{\@secondoftwo}%
\providecommand \bibfield  [0]{\@secondoftwo}%
\providecommand \translation [1]{[#1]}%
\providecommand \BibitemOpen [0]{}%
\providecommand \bibitemStop [0]{}%
\providecommand \bibitemNoStop [0]{.\EOS\space}%
\providecommand \EOS [0]{\spacefactor3000\relax}%
\providecommand \BibitemShut  [1]{\csname bibitem#1\endcsname}%
\let\auto@bib@innerbib\@empty
\bibitem [{\citenamefont {Agra\"it}, \citenamefont {Yeyati},\ and\
  \citenamefont {van Ruitenbeek}(2003)}]{Agrait2003}%
  \BibitemOpen
  \bibfield  {author} {\bibinfo {author} {\bibfnamefont {N.}~\bibnamefont
  {Agra\"it}}, \bibinfo {author} {\bibfnamefont {A.~L.}\ \bibnamefont
  {Yeyati}}, \ and\ \bibinfo {author} {\bibfnamefont {J.~M.}\ \bibnamefont {van
  Ruitenbeek}},\ }\href {\doibase 10.1016/S0370-1573(02)00633-6} {\bibfield
  {journal} {\bibinfo  {journal} {Phys. Rep.}\ }\textbf {\bibinfo {volume}
  {377}},\ \bibinfo {pages} {81} (\bibinfo {year} {2003})}\BibitemShut
  {NoStop}%
\bibitem [{\citenamefont {Venkataraman}\ \emph {et~al.}(2006)\citenamefont
  {Venkataraman}, \citenamefont {Klare}, \citenamefont {Nuckolls},
  \citenamefont {Hybertsen},\ and\ \citenamefont
  {Steigerwald}}]{Venkataraman2006}%
  \BibitemOpen
  \bibfield  {author} {\bibinfo {author} {\bibfnamefont {L.}~\bibnamefont
  {Venkataraman}}, \bibinfo {author} {\bibfnamefont {J.~E.}\ \bibnamefont
  {Klare}}, \bibinfo {author} {\bibfnamefont {C.}~\bibnamefont {Nuckolls}},
  \bibinfo {author} {\bibfnamefont {M.~S.}\ \bibnamefont {Hybertsen}}, \ and\
  \bibinfo {author} {\bibfnamefont {M.~L.}\ \bibnamefont {Steigerwald}},\
  }\href {\doibase 10.1038/nature05037} {\bibfield  {journal} {\bibinfo
  {journal} {Nature}\ }\textbf {\bibinfo {volume} {442}},\ \bibinfo {pages}
  {904} (\bibinfo {year} {2006})}\BibitemShut {NoStop}%
\bibitem [{\citenamefont {Osorio}\ \emph {et~al.}(2007)\citenamefont {Osorio},
  \citenamefont {O'Neill}, \citenamefont {Wegewijs}, \citenamefont
  {Stuhr-Hansen}, \citenamefont {Paaske}, \citenamefont {Bj{\o}rnholm},\ and\
  \citenamefont {van~der Zant}}]{Osorio2007}%
  \BibitemOpen
  \bibfield  {author} {\bibinfo {author} {\bibfnamefont {E.~A.}\ \bibnamefont
  {Osorio}}, \bibinfo {author} {\bibfnamefont {K.}~\bibnamefont {O'Neill}},
  \bibinfo {author} {\bibfnamefont {M.}~\bibnamefont {Wegewijs}}, \bibinfo
  {author} {\bibfnamefont {N.}~\bibnamefont {Stuhr-Hansen}}, \bibinfo {author}
  {\bibfnamefont {J.}~\bibnamefont {Paaske}}, \bibinfo {author} {\bibfnamefont
  {T.}~\bibnamefont {Bj{\o}rnholm}}, \ and\ \bibinfo {author} {\bibfnamefont
  {H.~S.~J.}\ \bibnamefont {van~der Zant}},\ }\href {\doibase
  10.1021/nl0715802} {\bibfield  {journal} {\bibinfo  {journal} {Nano Lett.}\
  }\textbf {\bibinfo {volume} {7}},\ \bibinfo {pages} {3336} (\bibinfo {year}
  {2007})}\BibitemShut {NoStop}%
\bibitem [{\citenamefont {Quek}\ \emph {et~al.}(2007)\citenamefont {Quek},
  \citenamefont {Venkataraman}, \citenamefont {Choi}, \citenamefont {Louie},
  \citenamefont {Hybertsen},\ and\ \citenamefont {Neaton}}]{Quek2007}%
  \BibitemOpen
  \bibfield  {author} {\bibinfo {author} {\bibfnamefont {S.~Y.}\ \bibnamefont
  {Quek}}, \bibinfo {author} {\bibfnamefont {L.}~\bibnamefont {Venkataraman}},
  \bibinfo {author} {\bibfnamefont {H.~J.}\ \bibnamefont {Choi}}, \bibinfo
  {author} {\bibfnamefont {S.~G.}\ \bibnamefont {Louie}}, \bibinfo {author}
  {\bibfnamefont {M.~S.}\ \bibnamefont {Hybertsen}}, \ and\ \bibinfo {author}
  {\bibfnamefont {J.~B.}\ \bibnamefont {Neaton}},\ }\href {\doibase
  10.1021/nl072058i} {\bibfield  {journal} {\bibinfo  {journal} {Nano Lett.}\
  }\textbf {\bibinfo {volume} {7}},\ \bibinfo {pages} {3477} (\bibinfo {year}
  {2007})}\BibitemShut {NoStop}%
\bibitem [{\citenamefont {Taylor}, \citenamefont {Guo},\ and\ \citenamefont
  {Wang}(2001)}]{Taylor2001}%
  \BibitemOpen
  \bibfield  {author} {\bibinfo {author} {\bibfnamefont {J.}~\bibnamefont
  {Taylor}}, \bibinfo {author} {\bibfnamefont {H.}~\bibnamefont {Guo}}, \ and\
  \bibinfo {author} {\bibfnamefont {J.}~\bibnamefont {Wang}},\ }\href {\doibase
  10.1103/PhysRevB.63.245407} {\bibfield  {journal} {\bibinfo  {journal} {Phys.
  Rev. B}\ }\textbf {\bibinfo {volume} {63}},\ \bibinfo {pages} {245407}
  (\bibinfo {year} {2001})}\BibitemShut {NoStop}%
\bibitem [{\citenamefont {Xue}, \citenamefont {Datta},\ and\ \citenamefont
  {Ratner}(2002)}]{Xue2002}%
  \BibitemOpen
  \bibfield  {author} {\bibinfo {author} {\bibfnamefont {Y.}~\bibnamefont
  {Xue}}, \bibinfo {author} {\bibfnamefont {S.}~\bibnamefont {Datta}}, \ and\
  \bibinfo {author} {\bibfnamefont {M.~A.}\ \bibnamefont {Ratner}},\ }\href
  {\doibase 10.1016/S0301-0104(02)00446-9} {\bibfield  {journal} {\bibinfo
  {journal} {Chem. Phys.}\ }\textbf {\bibinfo {volume} {281}},\ \bibinfo
  {pages} {151} (\bibinfo {year} {2002})}\BibitemShut {NoStop}%
\bibitem [{\citenamefont {Evers}, \citenamefont {Weigend},\ and\ \citenamefont
  {Koentopp}(2003)}]{Evers2003}%
  \BibitemOpen
  \bibfield  {author} {\bibinfo {author} {\bibfnamefont {F.}~\bibnamefont
  {Evers}}, \bibinfo {author} {\bibfnamefont {F.}~\bibnamefont {Weigend}}, \
  and\ \bibinfo {author} {\bibfnamefont {F.}~\bibnamefont {Koentopp}},\ }\href
  {\doibase 10.1016/S1386-9477(02)01006-8} {\bibfield  {journal} {\bibinfo
  {journal} {Physica E}\ }\textbf {\bibinfo {volume} {18}},\ \bibinfo {pages}
  {255} (\bibinfo {year} {2003})}\BibitemShut {NoStop}%
\bibitem [{\citenamefont {Qian}\ \emph {et~al.}(2007)\citenamefont {Qian},
  \citenamefont {Li}, \citenamefont {Hou}, \citenamefont {Xue},\ and\
  \citenamefont {Sanvito}}]{Qian2007}%
  \BibitemOpen
  \bibfield  {author} {\bibinfo {author} {\bibfnamefont {A.}~\bibnamefont
  {Qian}}, \bibinfo {author} {\bibfnamefont {R.}~\bibnamefont {Li}}, \bibinfo
  {author} {\bibfnamefont {S.}~\bibnamefont {Hou}}, \bibinfo {author}
  {\bibfnamefont {Z.}~\bibnamefont {Xue}}, \ and\ \bibinfo {author}
  {\bibfnamefont {S.}~\bibnamefont {Sanvito}},\ }\href {\doibase
  10.1063/1.2804876} {\bibfield  {journal} {\bibinfo  {journal} {J. Chem.
  Phys.}\ }\textbf {\bibinfo {volume} {127}},\ \bibinfo {pages} {194710}
  (\bibinfo {year} {2007})}\BibitemShut {NoStop}%
\bibitem [{\citenamefont {Brandbyge}\ \emph {et~al.}(2002)\citenamefont
  {Brandbyge}, \citenamefont {Mozos}, \citenamefont {Ordej\'on}, \citenamefont
  {Taylor},\ and\ \citenamefont {Stokbro}}]{Brandbyge2002}%
  \BibitemOpen
  \bibfield  {author} {\bibinfo {author} {\bibfnamefont {M.}~\bibnamefont
  {Brandbyge}}, \bibinfo {author} {\bibfnamefont {J.-L.}\ \bibnamefont
  {Mozos}}, \bibinfo {author} {\bibfnamefont {P.}~\bibnamefont {Ordej\'on}},
  \bibinfo {author} {\bibfnamefont {J.}~\bibnamefont {Taylor}}, \ and\ \bibinfo
  {author} {\bibfnamefont {K.}~\bibnamefont {Stokbro}},\ }\href {\doibase
  10.1103/PhysRevB.65.165401} {\bibfield  {journal} {\bibinfo  {journal} {Phys.
  Rev. B}\ }\textbf {\bibinfo {volume} {65}},\ \bibinfo {pages} {165401}
  (\bibinfo {year} {2002})}\BibitemShut {NoStop}%
\bibitem [{\citenamefont {Stokbro}\ \emph
  {et~al.}(2003{\natexlab{a}})\citenamefont {Stokbro}, \citenamefont {Taylor},
  \citenamefont {Brandbyge},\ and\ \citenamefont {Ordej\'on}}]{Stokbro2003b}%
  \BibitemOpen
  \bibfield  {author} {\bibinfo {author} {\bibfnamefont {K.}~\bibnamefont
  {Stokbro}}, \bibinfo {author} {\bibfnamefont {J.}~\bibnamefont {Taylor}},
  \bibinfo {author} {\bibfnamefont {M.}~\bibnamefont {Brandbyge}}, \ and\
  \bibinfo {author} {\bibfnamefont {P.}~\bibnamefont {Ordej\'on}},\ }\href
  {\doibase 10.1196/annals.1292.014} {\bibfield  {journal} {\bibinfo  {journal}
  {Ann. N.Y. Acad. Sci.}\ }\textbf {\bibinfo {volume} {1006}},\ \bibinfo
  {pages} {212} (\bibinfo {year} {2003}{\natexlab{a}})}\BibitemShut {NoStop}%
\bibitem [{\citenamefont {Rocha}\ \emph {et~al.}(2006)\citenamefont {Rocha},
  \citenamefont {Garc\'ia-Su\'arez}, \citenamefont {Bailey}, \citenamefont
  {Lambert}, \citenamefont {Ferrer},\ and\ \citenamefont
  {Sanvito}}]{Rocha2006}%
  \BibitemOpen
  \bibfield  {author} {\bibinfo {author} {\bibfnamefont {A.~R.}\ \bibnamefont
  {Rocha}}, \bibinfo {author} {\bibfnamefont {V.~M.}\ \bibnamefont
  {Garc\'ia-Su\'arez}}, \bibinfo {author} {\bibfnamefont {S.}~\bibnamefont
  {Bailey}}, \bibinfo {author} {\bibfnamefont {C.}~\bibnamefont {Lambert}},
  \bibinfo {author} {\bibfnamefont {J.}~\bibnamefont {Ferrer}}, \ and\ \bibinfo
  {author} {\bibfnamefont {S.}~\bibnamefont {Sanvito}},\ }\href {\doibase
  10.1103/PhysRevB.73.085414} {\bibfield  {journal} {\bibinfo  {journal} {Phys.
  Rev. B}\ }\textbf {\bibinfo {volume} {73}},\ \bibinfo {pages} {085414}
  (\bibinfo {year} {2006})}\BibitemShut {NoStop}%
\bibitem [{\citenamefont {Fonseca~Guerra}\ \emph {et~al.}(1998)\citenamefont
  {Fonseca~Guerra}, \citenamefont {Snijders}, \citenamefont {te~Velde},\ and\
  \citenamefont {Baerends}}]{FonsecaGuerra1998}%
  \BibitemOpen
  \bibfield  {author} {\bibinfo {author} {\bibfnamefont {C.}~\bibnamefont
  {Fonseca~Guerra}}, \bibinfo {author} {\bibfnamefont {J.~G.}\ \bibnamefont
  {Snijders}}, \bibinfo {author} {\bibfnamefont {G.}~\bibnamefont {te~Velde}},
  \ and\ \bibinfo {author} {\bibfnamefont {E.~J.}\ \bibnamefont {Baerends}},\
  }\href {\doibase 10.1007/s002140050353} {\bibfield  {journal} {\bibinfo
  {journal} {Theor. Chem. Acc.}\ }\textbf {\bibinfo {volume} {99}},\ \bibinfo
  {pages} {391} (\bibinfo {year} {1998})}\BibitemShut {NoStop}%
\bibitem [{\citenamefont {te~Velde}\ \emph {et~al.}(2001)\citenamefont
  {te~Velde}, \citenamefont {Bickelhaupt}, \citenamefont {van Gisbergen},
  \citenamefont {Fonseca~Guerra}, \citenamefont {Baerends}, \citenamefont
  {Snijders},\ and\ \citenamefont {Ziegler}}]{Velde2001}%
  \BibitemOpen
  \bibfield  {author} {\bibinfo {author} {\bibfnamefont {G.}~\bibnamefont
  {te~Velde}}, \bibinfo {author} {\bibfnamefont {F.~M.}\ \bibnamefont
  {Bickelhaupt}}, \bibinfo {author} {\bibfnamefont {S.~J.~A.}\ \bibnamefont
  {van Gisbergen}}, \bibinfo {author} {\bibfnamefont {C.}~\bibnamefont
  {Fonseca~Guerra}}, \bibinfo {author} {\bibfnamefont {E.~J.}\ \bibnamefont
  {Baerends}}, \bibinfo {author} {\bibfnamefont {J.~G.}\ \bibnamefont
  {Snijders}}, \ and\ \bibinfo {author} {\bibfnamefont {T.}~\bibnamefont
  {Ziegler}},\ }\href {\doibase 10.1002/jcc.1056} {\bibfield  {journal}
  {\bibinfo  {journal} {J. Comput. Chem.}\ }\textbf {\bibinfo {volume} {22}},\
  \bibinfo {pages} {931} (\bibinfo {year} {2001})}\BibitemShut {NoStop}%
\bibitem [{\citenamefont {te~Velde}\ and\ \citenamefont
  {Baerends}(1991)}]{Velde1991}%
  \BibitemOpen
  \bibfield  {author} {\bibinfo {author} {\bibfnamefont {G.}~\bibnamefont
  {te~Velde}}\ and\ \bibinfo {author} {\bibfnamefont {E.~J.}\ \bibnamefont
  {Baerends}},\ }\href {\doibase 10.1103/PhysRevB.44.7888} {\bibfield
  {journal} {\bibinfo  {journal} {Phys. Rev. B}\ }\textbf {\bibinfo {volume}
  {44}},\ \bibinfo {pages} {7888} (\bibinfo {year} {1991})}\BibitemShut
  {NoStop}%
\bibitem [{\citenamefont {Wiesenekker}\ and\ \citenamefont
  {Baerends}(1991)}]{Wiesenekker1991}%
  \BibitemOpen
  \bibfield  {author} {\bibinfo {author} {\bibfnamefont {G.}~\bibnamefont
  {Wiesenekker}}\ and\ \bibinfo {author} {\bibfnamefont {E.~J.}\ \bibnamefont
  {Baerends}},\ }\href {\doibase 10.1088/0953-8984/3/35/005} {\bibfield
  {journal} {\bibinfo  {journal} {J. Phys.: Condens. Matter}\ }\textbf
  {\bibinfo {volume} {3}},\ \bibinfo {pages} {6721} (\bibinfo {year}
  {1991})}\BibitemShut {NoStop}%
\bibitem [{\citenamefont {Verzijl}\ and\ \citenamefont
  {Thijssen}(2012)}]{Verzijl2012}%
  \BibitemOpen
  \bibfield  {author} {\bibinfo {author} {\bibfnamefont {C.~J.~O.}\
  \bibnamefont {Verzijl}}\ and\ \bibinfo {author} {\bibfnamefont {J.~M.}\
  \bibnamefont {Thijssen}},\ }\href {\doibase 10.1021/jp3044225} {\bibfield
  {journal} {\bibinfo  {journal} {J. Phys. Chem. C}\ }\textbf {\bibinfo
  {volume} {116}},\ \bibinfo {pages} {24393} (\bibinfo {year}
  {2012})}\BibitemShut {NoStop}%
\bibitem [{\citenamefont {Ernzerhof}, \citenamefont {Zhuang},\ and\
  \citenamefont {Rocheleau}(2005)}]{Ernzerhof2005}%
  \BibitemOpen
  \bibfield  {author} {\bibinfo {author} {\bibfnamefont {M.}~\bibnamefont
  {Ernzerhof}}, \bibinfo {author} {\bibfnamefont {M.}~\bibnamefont {Zhuang}}, \
  and\ \bibinfo {author} {\bibfnamefont {P.}~\bibnamefont {Rocheleau}},\ }\href
  {\doibase http://dx.doi.org/10.1063/1.2049249} {\bibfield  {journal}
  {\bibinfo  {journal} {J. Chem. Phys.}\ }\textbf {\bibinfo {volume} {123}},\
  \bibinfo {pages} {134704} (\bibinfo {year} {2005})}\BibitemShut {NoStop}%
\bibitem [{\citenamefont {Solomon}\ \emph {et~al.}(2008)\citenamefont
  {Solomon}, \citenamefont {Andrews}, \citenamefont {Hansen}, \citenamefont
  {Goldsmith}, \citenamefont {Wasielewski}, \citenamefont {Van~Duyne},\ and\
  \citenamefont {Ratner}}]{Solomon2008}%
  \BibitemOpen
  \bibfield  {author} {\bibinfo {author} {\bibfnamefont {G.~C.}\ \bibnamefont
  {Solomon}}, \bibinfo {author} {\bibfnamefont {D.~Q.}\ \bibnamefont
  {Andrews}}, \bibinfo {author} {\bibfnamefont {T.}~\bibnamefont {Hansen}},
  \bibinfo {author} {\bibfnamefont {R.~H.}\ \bibnamefont {Goldsmith}}, \bibinfo
  {author} {\bibfnamefont {M.~R.}\ \bibnamefont {Wasielewski}}, \bibinfo
  {author} {\bibfnamefont {R.~P.}\ \bibnamefont {Van~Duyne}}, \ and\ \bibinfo
  {author} {\bibfnamefont {M.~A.}\ \bibnamefont {Ratner}},\ }\href {\doibase
  http://dx.doi.org/10.1063/1.2958275} {\bibfield  {journal} {\bibinfo
  {journal} {J. Chem. Phys.}\ }\textbf {\bibinfo {volume} {129}},\ \bibinfo
  {pages} {054701} (\bibinfo {year} {2008})}\BibitemShut {NoStop}%
\bibitem [{\citenamefont {Rinc\'{o}n}\ \emph {et~al.}(2009)\citenamefont
  {Rinc\'{o}n}, \citenamefont {Hallberg}, \citenamefont {Aligia},\ and\
  \citenamefont {Ramasesha}}]{Rincon2009}%
  \BibitemOpen
  \bibfield  {author} {\bibinfo {author} {\bibfnamefont {J.}~\bibnamefont
  {Rinc\'{o}n}}, \bibinfo {author} {\bibfnamefont {K.}~\bibnamefont
  {Hallberg}}, \bibinfo {author} {\bibfnamefont {A.~A.}\ \bibnamefont
  {Aligia}}, \ and\ \bibinfo {author} {\bibfnamefont {S.}~\bibnamefont
  {Ramasesha}},\ }\href {\doibase 10.1103/PhysRevLett.103.266807} {\bibfield
  {journal} {\bibinfo  {journal} {Phys. Rev. Lett.}\ }\textbf {\bibinfo
  {volume} {103}},\ \bibinfo {pages} {266807} (\bibinfo {year}
  {2009})}\BibitemShut {NoStop}%
\bibitem [{\citenamefont {Markussen}, \citenamefont {Schi\"{o}tz},\ and\
  \citenamefont {Thygesen}(2010)}]{Markussen2010}%
  \BibitemOpen
  \bibfield  {author} {\bibinfo {author} {\bibfnamefont {T.}~\bibnamefont
  {Markussen}}, \bibinfo {author} {\bibfnamefont {J.}~\bibnamefont
  {Schi\"{o}tz}}, \ and\ \bibinfo {author} {\bibfnamefont {K.~S.}\ \bibnamefont
  {Thygesen}},\ }\href {\doibase http://dx.doi.org/10.1063/1.3451265}
  {\bibfield  {journal} {\bibinfo  {journal} {J. Chem. Phys.}\ }\textbf
  {\bibinfo {volume} {132}},\ \bibinfo {pages} {224104} (\bibinfo {year}
  {2010})}\BibitemShut {NoStop}%
\bibitem [{\citenamefont {Di~Ventra}, \citenamefont {Pantelides},\ and\
  \citenamefont {Lang}(2000)}]{DiVentra2000}%
  \BibitemOpen
  \bibfield  {author} {\bibinfo {author} {\bibfnamefont {M.}~\bibnamefont
  {Di~Ventra}}, \bibinfo {author} {\bibfnamefont {S.~T.}\ \bibnamefont
  {Pantelides}}, \ and\ \bibinfo {author} {\bibfnamefont {N.~D.}\ \bibnamefont
  {Lang}},\ }\href {\doibase 10.1103/PhysRevLett.84.979} {\bibfield  {journal}
  {\bibinfo  {journal} {Phys. Rev. Lett.}\ }\textbf {\bibinfo {volume} {84}},\
  \bibinfo {pages} {979} (\bibinfo {year} {2000})}\BibitemShut {NoStop}%
\bibitem [{\citenamefont {Di~Ventra}\ and\ \citenamefont
  {Lang}(2001)}]{DiVentra2001}%
  \BibitemOpen
  \bibfield  {author} {\bibinfo {author} {\bibfnamefont {M.}~\bibnamefont
  {Di~Ventra}}\ and\ \bibinfo {author} {\bibfnamefont {N.~D.}\ \bibnamefont
  {Lang}},\ }\href {\doibase 10.1103/PhysRevB.65.045402} {\bibfield  {journal}
  {\bibinfo  {journal} {Phys. Rev. B}\ }\textbf {\bibinfo {volume} {65}},\
  \bibinfo {pages} {045402} (\bibinfo {year} {2001})}\BibitemShut {NoStop}%
\bibitem [{\citenamefont {Ke}, \citenamefont {Baranger},\ and\ \citenamefont
  {Yang}(2004)}]{Ke2004}%
  \BibitemOpen
  \bibfield  {author} {\bibinfo {author} {\bibfnamefont {S.-H.}\ \bibnamefont
  {Ke}}, \bibinfo {author} {\bibfnamefont {H.~U.}\ \bibnamefont {Baranger}}, \
  and\ \bibinfo {author} {\bibfnamefont {W.}~\bibnamefont {Yang}},\ }\href
  {\doibase 10.1103/PhysRevB.70.085410} {\bibfield  {journal} {\bibinfo
  {journal} {Phys. Rev. B}\ }\textbf {\bibinfo {volume} {70}},\ \bibinfo
  {pages} {085410} (\bibinfo {year} {2004})}\BibitemShut {NoStop}%
\bibitem [{\citenamefont {Evers}\ and\ \citenamefont
  {Arnold}(2006)}]{Rothig2006}%
  \BibitemOpen
  \bibfield  {author} {\bibinfo {author} {\bibfnamefont {F.}~\bibnamefont
  {Evers}}\ and\ \bibinfo {author} {\bibfnamefont {A.}~\bibnamefont {Arnold}},\
  }\href@noop {} {\emph {\bibinfo {title} {CFN Lectures on Functional
  Nanostructures}}},\ edited by\ \bibinfo {editor} {\bibfnamefont
  {C.}~\bibnamefont {R\"othig}}, \bibinfo {editor} {\bibfnamefont
  {G.}~\bibnamefont {Sch\"on}}, \ and\ \bibinfo {editor} {\bibfnamefont
  {M.}~\bibnamefont {Vojta}},\ \bibinfo {series} {Springer Lecture Notes in
  Physics}, Vol.~\bibinfo {volume} {2}\ (\bibinfo  {publisher} {Springer},\
  \bibinfo {year} {2006})\ pp.\ \bibinfo {pages} {27--54}\BibitemShut {NoStop}%
\bibitem [{\citenamefont {Arnold}, \citenamefont {Weigend},\ and\ \citenamefont
  {Evers}(2007)}]{Arnold2007}%
  \BibitemOpen
  \bibfield  {author} {\bibinfo {author} {\bibfnamefont {A.}~\bibnamefont
  {Arnold}}, \bibinfo {author} {\bibfnamefont {F.}~\bibnamefont {Weigend}}, \
  and\ \bibinfo {author} {\bibfnamefont {F.}~\bibnamefont {Evers}},\ }\href
  {\doibase 10.1063/1.2716664} {\bibfield  {journal} {\bibinfo  {journal} {J.
  Chem. Phys.}\ }\textbf {\bibinfo {volume} {126}},\ \bibinfo {pages} {174101}
  (\bibinfo {year} {2007})}\BibitemShut {NoStop}%
\bibitem [{\citenamefont {Stokbro}\ \emph
  {et~al.}(2003{\natexlab{b}})\citenamefont {Stokbro}, \citenamefont {Taylor},
  \citenamefont {Brandbyge}, \citenamefont {Mozos},\ and\ \citenamefont
  {Ordej\'on}}]{Stokbro2003a}%
  \BibitemOpen
  \bibfield  {author} {\bibinfo {author} {\bibfnamefont {K.}~\bibnamefont
  {Stokbro}}, \bibinfo {author} {\bibfnamefont {J.}~\bibnamefont {Taylor}},
  \bibinfo {author} {\bibfnamefont {M.}~\bibnamefont {Brandbyge}}, \bibinfo
  {author} {\bibfnamefont {J.-L.}\ \bibnamefont {Mozos}}, \ and\ \bibinfo
  {author} {\bibfnamefont {P.}~\bibnamefont {Ordej\'on}},\ }\href {\doibase
  10.1016/S0927-0256(02)00439-1} {\bibfield  {journal} {\bibinfo  {journal}
  {Comp. Mat. Sci.}\ }\textbf {\bibinfo {volume} {27}},\ \bibinfo {pages} {151}
  (\bibinfo {year} {2003}{\natexlab{b}})}\BibitemShut {NoStop}%
\bibitem [{\citenamefont {Herrmann}\ and\ \citenamefont
  {Solomon}(2010)}]{ARTAIOS}%
  \BibitemOpen
  \bibfield  {author} {\bibinfo {author} {\bibfnamefont {C.}~\bibnamefont
  {Herrmann}}\ and\ \bibinfo {author} {\bibfnamefont {G.~C.}\ \bibnamefont
  {Solomon}},\ }\href@noop {} {\enquote {\bibinfo {title} {Artaios - a
  transport code for postprocessing quantum chemical electronic structure
  calculations},}\ }\bibinfo {type} {Tech. Rep.}\ (\bibinfo  {institution}
  {Northwestern University},\ \bibinfo {year} {2010})\BibitemShut {NoStop}%
\bibitem [{\citenamefont {Sen}\ and\ \citenamefont {Kaun}(2010)}]{Sen2010}%
  \BibitemOpen
  \bibfield  {author} {\bibinfo {author} {\bibfnamefont {A.}~\bibnamefont
  {Sen}}\ and\ \bibinfo {author} {\bibfnamefont {C.-C.}\ \bibnamefont {Kaun}},\
  }\href {\doibase 10.1021/nn101840a} {\bibfield  {journal} {\bibinfo
  {journal} {ACS Nano}\ }\textbf {\bibinfo {volume} {4}},\ \bibinfo {pages}
  {6404} (\bibinfo {year} {2010})}\BibitemShut {NoStop}%
\bibitem [{\citenamefont {Pontes}\ \emph {et~al.}(2011)\citenamefont {Pontes},
  \citenamefont {Rocha}, \citenamefont {Sanvito}, \citenamefont {Fazzio},\ and\
  \citenamefont {da~Silva}}]{Pontes2011}%
  \BibitemOpen
  \bibfield  {author} {\bibinfo {author} {\bibfnamefont {R.~B.}\ \bibnamefont
  {Pontes}}, \bibinfo {author} {\bibfnamefont {A.~R.}\ \bibnamefont {Rocha}},
  \bibinfo {author} {\bibfnamefont {S.}~\bibnamefont {Sanvito}}, \bibinfo
  {author} {\bibfnamefont {A.}~\bibnamefont {Fazzio}}, \ and\ \bibinfo {author}
  {\bibfnamefont {A.~J.~R.}\ \bibnamefont {da~Silva}},\ }\href {\doibase
  10.1021/nn101628w} {\bibfield  {journal} {\bibinfo  {journal} {ACS Nano}\
  }\textbf {\bibinfo {volume} {5}},\ \bibinfo {pages} {795} (\bibinfo {year}
  {2011})}\BibitemShut {NoStop}%
\bibitem [{\citenamefont {Ke}, \citenamefont {Baranger},\ and\ \citenamefont
  {Yang}(2005)}]{Ke2005}%
  \BibitemOpen
  \bibfield  {author} {\bibinfo {author} {\bibfnamefont {S.-H.}\ \bibnamefont
  {Ke}}, \bibinfo {author} {\bibfnamefont {H.~U.}\ \bibnamefont {Baranger}}, \
  and\ \bibinfo {author} {\bibfnamefont {W.}~\bibnamefont {Yang}},\ }\href
  {\doibase 10.1063/1.1851496} {\bibfield  {journal} {\bibinfo  {journal} {J.
  Chem. Phys.}\ }\textbf {\bibinfo {volume} {122}},\ \bibinfo {eid} {074704}
  (\bibinfo {year} {2005})}\BibitemShut {NoStop}%
\bibitem [{\citenamefont {Ke}, \citenamefont {Baranger},\ and\ \citenamefont
  {Yang}(2007{\natexlab{a}})}]{Ke2007a}%
  \BibitemOpen
  \bibfield  {author} {\bibinfo {author} {\bibfnamefont {S.-H.}\ \bibnamefont
  {Ke}}, \bibinfo {author} {\bibfnamefont {H.~U.}\ \bibnamefont {Baranger}}, \
  and\ \bibinfo {author} {\bibfnamefont {W.}~\bibnamefont {Yang}},\ }\href
  {\doibase 10.1063/1.2770718} {\bibfield  {journal} {\bibinfo  {journal} {J.
  Chem. Phys.}\ }\textbf {\bibinfo {volume} {127}},\ \bibinfo {eid} {144107}
  (\bibinfo {year} {2007}{\natexlab{a}})}\BibitemShut {NoStop}%
\bibitem [{\citenamefont {Ke}, \citenamefont {Baranger},\ and\ \citenamefont
  {Yang}(2007{\natexlab{b}})}]{Ke2007b}%
  \BibitemOpen
  \bibfield  {author} {\bibinfo {author} {\bibfnamefont {S.-H.}\ \bibnamefont
  {Ke}}, \bibinfo {author} {\bibfnamefont {H.~U.}\ \bibnamefont {Baranger}}, \
  and\ \bibinfo {author} {\bibfnamefont {W.}~\bibnamefont {Yang}},\ }\href
  {\doibase 10.1063/1.2743004} {\bibfield  {journal} {\bibinfo  {journal} {J.
  Chem. Phys.}\ }\textbf {\bibinfo {volume} {126}},\ \bibinfo {eid} {201102}
  (\bibinfo {year} {2007}{\natexlab{b}})}\BibitemShut {NoStop}%
\bibitem [{\citenamefont {Hoft}, \citenamefont {Ford},\ and\ \citenamefont
  {Cortie}(2007)}]{Hoft2007}%
  \BibitemOpen
  \bibfield  {author} {\bibinfo {author} {\bibfnamefont {R.~C.}\ \bibnamefont
  {Hoft}}, \bibinfo {author} {\bibfnamefont {M.~J.}\ \bibnamefont {Ford}}, \
  and\ \bibinfo {author} {\bibfnamefont {M.~B.}\ \bibnamefont {Cortie}},\
  }\href {\doibase 10.1080/08927020701435811} {\bibfield  {journal} {\bibinfo
  {journal} {Mol. Simul.}\ }\textbf {\bibinfo {volume} {33}},\ \bibinfo {pages}
  {897} (\bibinfo {year} {2007})}\BibitemShut {NoStop}%
\bibitem [{\citenamefont {Guo}\ \emph {et~al.}(2006)\citenamefont {Guo},
  \citenamefont {Small}, \citenamefont {Klare}, \citenamefont {Wang},
  \citenamefont {Purewal}, \citenamefont {Tam}, \citenamefont {Hong},
  \citenamefont {Caldwell}, \citenamefont {Huang}, \citenamefont {O'Brien},
  \citenamefont {Yan}, \citenamefont {Breslow}, \citenamefont {Wind},
  \citenamefont {Hone}, \citenamefont {Kim},\ and\ \citenamefont
  {Nuckolls}}]{Guo2006}%
  \BibitemOpen
  \bibfield  {author} {\bibinfo {author} {\bibfnamefont {X.}~\bibnamefont
  {Guo}}, \bibinfo {author} {\bibfnamefont {J.~P.}\ \bibnamefont {Small}},
  \bibinfo {author} {\bibfnamefont {J.~E.}\ \bibnamefont {Klare}}, \bibinfo
  {author} {\bibfnamefont {Y.}~\bibnamefont {Wang}}, \bibinfo {author}
  {\bibfnamefont {M.~S.}\ \bibnamefont {Purewal}}, \bibinfo {author}
  {\bibfnamefont {I.~W.}\ \bibnamefont {Tam}}, \bibinfo {author} {\bibfnamefont
  {B.~H.}\ \bibnamefont {Hong}}, \bibinfo {author} {\bibfnamefont
  {R.}~\bibnamefont {Caldwell}}, \bibinfo {author} {\bibfnamefont
  {L.}~\bibnamefont {Huang}}, \bibinfo {author} {\bibfnamefont
  {S.}~\bibnamefont {O'Brien}}, \bibinfo {author} {\bibfnamefont
  {J.}~\bibnamefont {Yan}}, \bibinfo {author} {\bibfnamefont {R.}~\bibnamefont
  {Breslow}}, \bibinfo {author} {\bibfnamefont {S.~J.}\ \bibnamefont {Wind}},
  \bibinfo {author} {\bibfnamefont {J.}~\bibnamefont {Hone}}, \bibinfo {author}
  {\bibfnamefont {P.}~\bibnamefont {Kim}}, \ and\ \bibinfo {author}
  {\bibfnamefont {C.}~\bibnamefont {Nuckolls}},\ }\href {\doibase
  10.1126/science.1120986} {\bibfield  {journal} {\bibinfo  {journal}
  {Science}\ }\textbf {\bibinfo {volume} {311}},\ \bibinfo {pages} {356}
  (\bibinfo {year} {2006})}\BibitemShut {NoStop}%
\bibitem [{\citenamefont {Feldman}\ \emph {et~al.}(2008)\citenamefont
  {Feldman}, \citenamefont {Steigerwald}, \citenamefont {Guo},\ and\
  \citenamefont {Nuckolls}}]{Feldman2008}%
  \BibitemOpen
  \bibfield  {author} {\bibinfo {author} {\bibfnamefont {A.~K.}\ \bibnamefont
  {Feldman}}, \bibinfo {author} {\bibfnamefont {M.~L.}\ \bibnamefont
  {Steigerwald}}, \bibinfo {author} {\bibfnamefont {X.}~\bibnamefont {Guo}}, \
  and\ \bibinfo {author} {\bibfnamefont {C.}~\bibnamefont {Nuckolls}},\ }\href
  {\doibase 10.1021/ar8000266} {\bibfield  {journal} {\bibinfo  {journal} {Acc.
  Chem. Res.}\ }\textbf {\bibinfo {volume} {41}},\ \bibinfo {pages} {1731}
  (\bibinfo {year} {2008})}\BibitemShut {NoStop}%
\bibitem [{\citenamefont {Prins}\ \emph {et~al.}(2011)\citenamefont {Prins},
  \citenamefont {Barreiro}, \citenamefont {Ruitenberg}, \citenamefont
  {Seldenthuis}, \citenamefont {Aliaga-Alcalde}, \citenamefont {Vandersypen},\
  and\ \citenamefont {van~der Zant}}]{Prins2011}%
  \BibitemOpen
  \bibfield  {author} {\bibinfo {author} {\bibfnamefont {F.}~\bibnamefont
  {Prins}}, \bibinfo {author} {\bibfnamefont {A.}~\bibnamefont {Barreiro}},
  \bibinfo {author} {\bibfnamefont {J.~W.}\ \bibnamefont {Ruitenberg}},
  \bibinfo {author} {\bibfnamefont {J.~S.}\ \bibnamefont {Seldenthuis}},
  \bibinfo {author} {\bibfnamefont {N.}~\bibnamefont {Aliaga-Alcalde}},
  \bibinfo {author} {\bibfnamefont {L.~M.~K.}\ \bibnamefont {Vandersypen}}, \
  and\ \bibinfo {author} {\bibfnamefont {H.~S.~J.}\ \bibnamefont {van~der
  Zant}},\ }\href {\doibase 10.1021/nl202065x} {\bibfield  {journal} {\bibinfo
  {journal} {Nano Lett.}\ }\textbf {\bibinfo {volume} {11}},\ \bibinfo {pages}
  {4607} (\bibinfo {year} {2011})}\BibitemShut {NoStop}%
\bibitem [{\citenamefont {Koentopp}, \citenamefont {Burke},\ and\ \citenamefont
  {Evers}(2006)}]{Koentopp2006}%
  \BibitemOpen
  \bibfield  {author} {\bibinfo {author} {\bibfnamefont {M.}~\bibnamefont
  {Koentopp}}, \bibinfo {author} {\bibfnamefont {K.}~\bibnamefont {Burke}}, \
  and\ \bibinfo {author} {\bibfnamefont {F.}~\bibnamefont {Evers}},\ }\href
  {\doibase 10.1103/PhysRevB.73.121403} {\bibfield  {journal} {\bibinfo
  {journal} {Phys. Rev. B}\ }\textbf {\bibinfo {volume} {73}},\ \bibinfo {eid}
  {121403} (\bibinfo {year} {2006})}\BibitemShut {NoStop}%
\bibitem [{\citenamefont {Perdew}\ and\ \citenamefont
  {Zunger}(1981)}]{Perdew1981}%
  \BibitemOpen
  \bibfield  {author} {\bibinfo {author} {\bibfnamefont {J.~P.}\ \bibnamefont
  {Perdew}}\ and\ \bibinfo {author} {\bibfnamefont {A.}~\bibnamefont
  {Zunger}},\ }\href {\doibase 10.1103/PhysRevB.23.5048} {\bibfield  {journal}
  {\bibinfo  {journal} {Phys. Rev. B}\ }\textbf {\bibinfo {volume} {23}},\
  \bibinfo {pages} {5048} (\bibinfo {year} {1981})}\BibitemShut {NoStop}%
\bibitem [{\citenamefont {Toher}\ \emph {et~al.}(2005)\citenamefont {Toher},
  \citenamefont {Filippetti}, \citenamefont {Sanvito},\ and\ \citenamefont
  {Burke}}]{Toher2005}%
  \BibitemOpen
  \bibfield  {author} {\bibinfo {author} {\bibfnamefont {C.}~\bibnamefont
  {Toher}}, \bibinfo {author} {\bibfnamefont {A.}~\bibnamefont {Filippetti}},
  \bibinfo {author} {\bibfnamefont {S.}~\bibnamefont {Sanvito}}, \ and\
  \bibinfo {author} {\bibfnamefont {K.}~\bibnamefont {Burke}},\ }\href
  {\doibase 10.1103/PhysRevLett.95.146402} {\bibfield  {journal} {\bibinfo
  {journal} {Phys. Rev. Lett.}\ }\textbf {\bibinfo {volume} {95}},\ \bibinfo
  {eid} {146402} (\bibinfo {year} {2005})}\BibitemShut {NoStop}%
\bibitem [{\citenamefont {Aryasetiawan}\ and\ \citenamefont
  {Gunnarsson}(1998)}]{Aryasetiawan1998}%
  \BibitemOpen
  \bibfield  {author} {\bibinfo {author} {\bibfnamefont {F.}~\bibnamefont
  {Aryasetiawan}}\ and\ \bibinfo {author} {\bibfnamefont {O.}~\bibnamefont
  {Gunnarsson}},\ }\href {\doibase 10.1088/0034-4885/61/3/002} {\bibfield
  {journal} {\bibinfo  {journal} {Rep. Prog. Phys.}\ }\textbf {\bibinfo
  {volume} {61}},\ \bibinfo {pages} {237} (\bibinfo {year} {1998})}\BibitemShut
  {NoStop}%
\bibitem [{\citenamefont {Thygesen}\ and\ \citenamefont
  {Rubio}(2009)}]{Thygesen2009}%
  \BibitemOpen
  \bibfield  {author} {\bibinfo {author} {\bibfnamefont {K.~S.}\ \bibnamefont
  {Thygesen}}\ and\ \bibinfo {author} {\bibfnamefont {A.}~\bibnamefont
  {Rubio}},\ }\href {\doibase 10.1103/PhysRevLett.102.046802} {\bibfield
  {journal} {\bibinfo  {journal} {Phys. Rev. Lett.}\ }\textbf {\bibinfo
  {volume} {102}},\ \bibinfo {eid} {046802} (\bibinfo {year}
  {2009})}\BibitemShut {NoStop}%
\bibitem [{\citenamefont {Jauho}, \citenamefont {Wingreen},\ and\ \citenamefont
  {Meir}(1994)}]{Jauho1994}%
  \BibitemOpen
  \bibfield  {author} {\bibinfo {author} {\bibfnamefont {A.-P.}\ \bibnamefont
  {Jauho}}, \bibinfo {author} {\bibfnamefont {N.~S.}\ \bibnamefont {Wingreen}},
  \ and\ \bibinfo {author} {\bibfnamefont {Y.}~\bibnamefont {Meir}},\ }\href
  {\doibase 10.1103/PhysRevB.50.5528} {\bibfield  {journal} {\bibinfo
  {journal} {Phys. Rev. B}\ }\textbf {\bibinfo {volume} {50}},\ \bibinfo
  {pages} {5528} (\bibinfo {year} {1994})}\BibitemShut {NoStop}%
\bibitem [{\citenamefont {Haug}\ and\ \citenamefont {Jauho}(1997)}]{Haug1997}%
  \BibitemOpen
  \bibfield  {author} {\bibinfo {author} {\bibfnamefont {H.}~\bibnamefont
  {Haug}}\ and\ \bibinfo {author} {\bibfnamefont {A.-P.}\ \bibnamefont
  {Jauho}},\ }\href@noop {} {\emph {\bibinfo {title} {Quantum Kinetics in
  Transport and Optics of Semi-conductors}}}\ (\bibinfo  {publisher}
  {Springer-Verlag},\ \bibinfo {year} {1997})\BibitemShut {NoStop}%
\bibitem [{\citenamefont {Newns}(1969)}]{Newns1969}%
  \BibitemOpen
  \bibfield  {author} {\bibinfo {author} {\bibfnamefont {D.~M.}\ \bibnamefont
  {Newns}},\ }\href {\doibase 10.1103/PhysRev.178.1123} {\bibfield  {journal}
  {\bibinfo  {journal} {Phys. Rev.}\ }\textbf {\bibinfo {volume} {178}},\
  \bibinfo {pages} {1123} (\bibinfo {year} {1969})}\BibitemShut {NoStop}%
\bibitem [{\citenamefont {Anderson}(1961)}]{Anderson1961}%
  \BibitemOpen
  \bibfield  {author} {\bibinfo {author} {\bibfnamefont {P.~W.}\ \bibnamefont
  {Anderson}},\ }\href {\doibase 10.1103/PhysRev.124.41} {\bibfield  {journal}
  {\bibinfo  {journal} {Phys. Rev.}\ }\textbf {\bibinfo {volume} {124}},\
  \bibinfo {pages} {41} (\bibinfo {year} {1961})}\BibitemShut {NoStop}%
\bibitem [{\citenamefont {Meir}\ and\ \citenamefont
  {Wingreen}(1992)}]{Meir1992}%
  \BibitemOpen
  \bibfield  {author} {\bibinfo {author} {\bibfnamefont {Y.}~\bibnamefont
  {Meir}}\ and\ \bibinfo {author} {\bibfnamefont {N.~S.}\ \bibnamefont
  {Wingreen}},\ }\href {\doibase 10.1103/PhysRevLett.68.2512} {\bibfield
  {journal} {\bibinfo  {journal} {Phys. Rev. Lett.}\ }\textbf {\bibinfo
  {volume} {68}},\ \bibinfo {pages} {2512} (\bibinfo {year}
  {1992})}\BibitemShut {NoStop}%
\bibitem [{\citenamefont {Smith}\ \emph {et~al.}(1974)\citenamefont {Smith},
  \citenamefont {Wertheim}, \citenamefont {H\"ufner},\ and\ \citenamefont
  {Traum}}]{Smith1974}%
  \BibitemOpen
  \bibfield  {author} {\bibinfo {author} {\bibfnamefont {N.~V.}\ \bibnamefont
  {Smith}}, \bibinfo {author} {\bibfnamefont {G.~K.}\ \bibnamefont {Wertheim}},
  \bibinfo {author} {\bibfnamefont {S.}~\bibnamefont {H\"ufner}}, \ and\
  \bibinfo {author} {\bibfnamefont {M.~M.}\ \bibnamefont {Traum}},\ }\href
  {\doibase 10.1103/PhysRevB.10.3197} {\bibfield  {journal} {\bibinfo
  {journal} {Phys. Rev. B}\ }\textbf {\bibinfo {volume} {10}},\ \bibinfo
  {pages} {3197} (\bibinfo {year} {1974})}\BibitemShut {NoStop}%
\bibitem [{\citenamefont {Jepsen}, \citenamefont {Gl\"otzel},\ and\
  \citenamefont {Mackintosh}(1981)}]{Jepsen1981}%
  \BibitemOpen
  \bibfield  {author} {\bibinfo {author} {\bibfnamefont {O.}~\bibnamefont
  {Jepsen}}, \bibinfo {author} {\bibfnamefont {D.}~\bibnamefont {Gl\"otzel}}, \
  and\ \bibinfo {author} {\bibfnamefont {A.~R.}\ \bibnamefont {Mackintosh}},\
  }\href {\doibase 10.1103/PhysRevB.23.2684} {\bibfield  {journal} {\bibinfo
  {journal} {Phys. Rev. B}\ }\textbf {\bibinfo {volume} {23}},\ \bibinfo
  {pages} {2684} (\bibinfo {year} {1981})}\BibitemShut {NoStop}%
\bibitem [{Note1()}]{Note1}%
  \BibitemOpen
  \bibinfo {note} {Technically, this only holds for an orthonormal basis, but
  it is always possible to transform to such a basis using L{\"o}wdin
  orthogonalization\cite {Lowdin1950}}\BibitemShut {NoStop}%
\bibitem [{\citenamefont {Kondo}\ \emph {et~al.}(2006)\citenamefont {Kondo},
  \citenamefont {Kino}, \citenamefont {Nara}, \citenamefont {Ozaki},\ and\
  \citenamefont {Ohno}}]{Kondo2006}%
  \BibitemOpen
  \bibfield  {author} {\bibinfo {author} {\bibfnamefont {H.}~\bibnamefont
  {Kondo}}, \bibinfo {author} {\bibfnamefont {H.}~\bibnamefont {Kino}},
  \bibinfo {author} {\bibfnamefont {J.}~\bibnamefont {Nara}}, \bibinfo {author}
  {\bibfnamefont {T.}~\bibnamefont {Ozaki}}, \ and\ \bibinfo {author}
  {\bibfnamefont {T.}~\bibnamefont {Ohno}},\ }\href {\doibase
  10.1103/PhysRevB.73.235323} {\bibfield  {journal} {\bibinfo  {journal} {Phys.
  Rev. B}\ }\textbf {\bibinfo {volume} {73}},\ \bibinfo {pages} {235323}
  (\bibinfo {year} {2006})}\BibitemShut {NoStop}%
\bibitem [{\citenamefont {Evers}, \citenamefont {Weigend},\ and\ \citenamefont
  {Koentopp}(2004)}]{Evers2004}%
  \BibitemOpen
  \bibfield  {author} {\bibinfo {author} {\bibfnamefont {F.}~\bibnamefont
  {Evers}}, \bibinfo {author} {\bibfnamefont {F.}~\bibnamefont {Weigend}}, \
  and\ \bibinfo {author} {\bibfnamefont {M.}~\bibnamefont {Koentopp}},\ }\href
  {\doibase 10.1103/PhysRevB.69.235411} {\bibfield  {journal} {\bibinfo
  {journal} {Phys. Rev. B}\ }\textbf {\bibinfo {volume} {69}},\ \bibinfo
  {pages} {235411} (\bibinfo {year} {2004})}\BibitemShut {NoStop}%
\bibitem [{Note2()}]{Note2}%
  \BibitemOpen
  \bibinfo {note} {For one or two layers the gold in the extended molecule is
  not sufficiently bulk-like to attain the correct Fermi energy, and large
  shifts are necessary to make the transmissions overlap.}\BibitemShut {Stop}%
\bibitem [{\citenamefont {Xue}\ and\ \citenamefont {Ratner}(2003)}]{Xue2003}%
  \BibitemOpen
  \bibfield  {author} {\bibinfo {author} {\bibfnamefont {Y.}~\bibnamefont
  {Xue}}\ and\ \bibinfo {author} {\bibfnamefont {M.~A.}\ \bibnamefont
  {Ratner}},\ }\href {\doibase 10.1103/PhysRevB.68.115406} {\bibfield
  {journal} {\bibinfo  {journal} {Phys. Rev. B}\ }\textbf {\bibinfo {volume}
  {68}},\ \bibinfo {pages} {115406} (\bibinfo {year} {2003})}\BibitemShut
  {NoStop}%
\bibitem [{\citenamefont {Ashcroft}\ and\ \citenamefont
  {Mermin}(1976)}]{Ashcroft1976}%
  \BibitemOpen
  \bibfield  {author} {\bibinfo {author} {\bibfnamefont {N.~W.}\ \bibnamefont
  {Ashcroft}}\ and\ \bibinfo {author} {\bibfnamefont {N.~D.}\ \bibnamefont
  {Mermin}},\ }\href@noop {} {\emph {\bibinfo {title} {Solid State Physics}}}\
  (\bibinfo  {publisher} {Brooks/Cole},\ \bibinfo {year} {1976})\BibitemShut
  {NoStop}%
\bibitem [{Note3()}]{Note3}%
  \BibitemOpen
  \bibinfo {note} {Using the PBE GGA potential shifts the transmission
  slightly, but otherwise leaves the spectrum unchanged.}\BibitemShut {Stop}%
\bibitem [{\citenamefont {Ning}\ \emph {et~al.}(2007)\citenamefont {Ning},
  \citenamefont {Li}, \citenamefont {Shen}, \citenamefont {Qian}, \citenamefont
  {Hou}, \citenamefont {Rocha},\ and\ \citenamefont {Sanvito}}]{Ning2007}%
  \BibitemOpen
  \bibfield  {author} {\bibinfo {author} {\bibfnamefont {J.}~\bibnamefont
  {Ning}}, \bibinfo {author} {\bibfnamefont {R.}~\bibnamefont {Li}}, \bibinfo
  {author} {\bibfnamefont {X.}~\bibnamefont {Shen}}, \bibinfo {author}
  {\bibfnamefont {Z.}~\bibnamefont {Qian}}, \bibinfo {author} {\bibfnamefont
  {S.}~\bibnamefont {Hou}}, \bibinfo {author} {\bibfnamefont {A.~R.}\
  \bibnamefont {Rocha}}, \ and\ \bibinfo {author} {\bibfnamefont
  {S.}~\bibnamefont {Sanvito}},\ }\href {\doibase
  10.1088/0957-4484/18/34/345203} {\bibfield  {journal} {\bibinfo  {journal}
  {Nanotechnology}\ }\textbf {\bibinfo {volume} {18}},\ \bibinfo {pages}
  {345203} (\bibinfo {year} {2007})}\BibitemShut {NoStop}%
\bibitem [{\citenamefont {L\"owdin}(1950)}]{Lowdin1950}%
  \BibitemOpen
  \bibfield  {author} {\bibinfo {author} {\bibfnamefont {P.-O.}\ \bibnamefont
  {L\"owdin}},\ }\href {\doibase 10.1063/1.1747632} {\bibfield  {journal}
  {\bibinfo  {journal} {J. Chem. Phys.}\ }\textbf {\bibinfo {volume} {18}},\
  \bibinfo {pages} {365} (\bibinfo {year} {1950})}\BibitemShut {NoStop}%
\end{thebibliography}
\end{document}